\documentclass[10pt,journal,compsoc]{IEEEtran}
\IEEEoverridecommandlockouts
\usepackage{url}
\usepackage{verbatim}
\usepackage{cite}
\usepackage{amsmath,amssymb,amsfonts}
\usepackage{lipsum}
\usepackage{graphicx}
\usepackage{textcomp}
\usepackage{xcolor}
\usepackage{subfigure}
\usepackage{amsthm}
\usepackage{dsfont}
\usepackage{algorithmic}
\usepackage{algorithm}

\newtheorem{theorem}{Theorem}
\newtheorem{corollary}{Corollary} 
\newtheorem{lemma}{Lemma}

\newtheorem{assum}{Assumption}

\def\BibTeX{{\rm B\kern-.05em{\sc i\kern-.025em b}\kern-.08em
    T\kern-.1667em\lower.7ex\hbox{E}\kern-.125emX}}
\begin{document}

\title{
Straggler Mitigation and Latency Optimization in Blockchain-based Hierarchical Federated Learning

\author{Zhilin Wang, Qin~Hu, Minghui Xu, Zehui Xiong}
\thanks{This work is partly supported by the US NSF under grant CNS-2105004.}
\IEEEcompsocitemizethanks{
\IEEEcompsocthanksitem Zhilin Wang and Qin Hu (corresponding author) are with the Department of Computer and Information Science, Indiana University-Purdue University Indianapolis, IN, 46202, USA. E-mail: \{wangzhil,qinhu\}@iu.edu
\IEEEcompsocthanksitem Minghui Xu is with the School of Computer Science and Technology, Shandong University, China. E-mail: mhxu@sdu.edu.cn
\IEEEcompsocthanksitem Zehui Xiong is with Pillar of Information Systems Technology and Design,
Singapore University of Technology Design, Singapore. E-mail: zehui\_xiong@sutd.edu.sg
}
}

\IEEEpubidadjcol

\IEEEtitleabstractindextext{

\begin{abstract}

Cloud-edge-device hierarchical federated learning (HFL) has been recently proposed to achieve communication-efficient and privacy-preserving distributed learning. However, there exist several critical challenges, such as the single point of failure and potential stragglers in both edge servers and local devices. To resolve these issues, we propose a decentralized and straggler-tolerant blockchain-based HFL (BHFL) framework. Specifically, a Raft-based consortium blockchain is deployed on edge servers to provide a distributed and trusted computing environment for global model aggregation in BHFL. To mitigate the influence of stragglers on learning, we propose a novel aggregation method, \textit{HieAvg}, which utilizes the historical weights of stragglers to estimate the missing submissions. Furthermore, we optimize the overall latency of BHFL by jointly considering the constraints of global model convergence and blockchain consensus delay. Theoretical analysis and experimental evaluation show that our proposed BHFL based on HieAvg can converge in the presence of stragglers, which performs better than the traditional methods even when the loss function is non-convex and the data on local devices are non-independent and identically distributed (non-IID).
\end{abstract}

}

\maketitle
\section{Introduction}
As a representative paradigm of distributed machine learning, federated learning (FL) significantly reduces the cost of data transmission and protects data privacy\cite{mcmahan2017communication, kairouz2021advances, li2020federated}. In FL, local devices (i.e., FL clients) upload the trained local models to the parameter server (i.e., aggregator) for global model aggregation. 
However, 
FL needs multiple rounds of global model aggregation to obtain the optimal model, which not only consumes substantial communication resources of local devices but may cause network congestion and thus long latency of receiving updates at the server. Therefore, communication efficiency becomes one of the major bottlenecks of FL.

Hierarchical federated learning (HFL) provides a promising solution to the above challenge\cite{abad2020hierarchical,lim2021decentralized,wainakh2020enhancing,wang2022demystifying}. The basic idea is to conduct multiple intermediate aggregations at proxy servers (e.g., edge servers) before global aggregation on the central server. 
Local devices upload the model updates to a closer proxy server for model aggregation, reducing their communication cost. As demonstrated in \cite{liu2020client}, HFL can effectively reduce communication latency. Nevertheless, HFL still faces many problems. First, HFL requires a central server for global model aggregation; once this central server is failed, HFL cannot work anymore. In addition, proxy servers in the intermediate layer introduce a new attack surface, where the privacy leakage of local model updates and other malicious attacks (e.g., model poisoning attacks) become severe threats\cite{wang2019beyond,fang2020local}. 

Blockchain \cite{nakamoto2008bitcoin, zheng2018blockchain}, as a distributed ledger technology, has been widely applied to the fields of distributed machine learning\cite{kim2018device,nguyen2021federated}. %
Considering that blockchain can establish a decentralized and trustless computing environment, we can similarly implement blockchain on proxy servers to take the place of the central server in HFL so as to reduce the risks of the single point of failure and malicious attacks. 
There exist some studies \cite{xu2022mudfl,zhang2021bc} deploying blockchain in HFL to protect the privacy and improve efficiency, 
termed blockchain-based HFL (BHFL); but they still rely on the central server to aggregate global models. 
Furthermore, applying blockchain on HFL can lead to extra latency during broadcasting, verification, and consensus to generate a new block. 
Although there is some research that optimizes the latency of BHFL by designing resource allocation mechanisms among devices \cite{nguyen2022latency}, it still cannot resolve the influence of blockchain consensus on latency.

Apart from the latency issue, the challenge of stragglers still remains unaddressed in BHFL. Here the straggler refers to any participant, including local devices and proxy servers, that cannot submit the model updates in time due to insufficient computing resources or unstable network conditions. 
The communication efficiency would be directly affected by the stragglers, as waiting for updates from all clients can lead to significant time consumption. Besides, in the case of permanent stragglers that never rejoin FL, simply abandoning their updates can lead to poor performance of the global model, especially when the data of local devices are non-independent and identically distributed (non-IID) \cite{zhao2018federated}.

The existing studies mitigating the impact of stragglers on FL usually utilize the \textit{coded computing technology} \cite{dhakal2019coded, prakash2020coded, schlegel2021codedpaddedfl} or manipulate \textit{delayed gradients} \cite{li2021stragglers,xu2021live}. The coded computing method requires extra encoding/decoding processes and data transmission, which is not computing or communication efficient to be applied to various deep learning models. The scheme of manipulating delayed gradients can well address stragglers lacking computing power, where they can still submit partial gradients for model aggregation; however, this method cannot deal with stragglers caused by network disconnection or congestion, where the aggregator cannot obtain any data from stragglers within the required time.
In addition to local devices, proxy servers can also become stragglers in HFL due to unexpected connection failures, which cannot be resolved by either of the above methods since proxy servers are not responsible for model training. 
Some research \cite{gu2021fast} utilizes the historical updates to predict the missing updates but the estimation bias could be significant.


To fill these gaps, we propose a fully decentralized BHFL framework, which is proven to be convergence-guaranteed even with stragglers existing in local devices and edge servers. Specifically, we deploy a lightweight Raft-based consortium blockchain \cite{huang2019performance} on edge servers to provide a secure and trusted computing environment for HFL.
To address the challenge of stragglers in BHFL, we design a novel aggregation algorithm, named \textit{hierarchical averaging (HieAvg)}, to aggregate model updates submitted from local devices and edge servers at the edge aggregation and global aggregation phases, respectively. The basic idea of HieAvg is to estimate the missing weights with the differences between the historical weights of stragglers, and HieAvg can work well with non-IID data and non-convex loss functions.
Further, to improve the system efficiency of BHFL, we optimize the overall latency of BHFL by balancing the performance of the global model and the time cost of blockchain consensus.


To the best of our knowledge, this is the first work to solve the problem of stragglers in BHFL. The proposed HieAvg is applicable to not only our considered BHFL framework but also the general HFL scenarios. The main contributions can be summarized below:
\begin{itemize}
    \item We propose a decentralized BHFL framework that can converge even when there are stragglers in both local devices and edge servers with non-convex loss function and non-IID data.
    
    \item We design HieAvg, a novel model aggregation method for BHFL, to mitigate the negative impact of stragglers by utilizing their historical weights to estimate the missing weights and its convergence is theoretically proved.
    
    \item We optimize the total latency of BHFL by deriving the optimal number of aggregation rounds on edge servers under the constraints of blockchain consensus time consumption and global model convergence.
    
    \item Rigorous theoretical analysis and extensive experiments are conducted to prove the convergence of BHFL with HieAvg and evaluate the validity and efficiency of our proposed schemes.
\end{itemize}

The rest of this paper is organized as below. We introduce the system model in Section \ref{system}. Then, we analyze the convergence of BHFL with HieAvg in Section \ref{convg}, and the latency optimization is shown in Section \ref{opt}. Next, we conduct experiments to support our framework and mechanisms in Section \ref{exp}. The related work is discussed in Section \ref{related}. Finally, we conclude this work in Section \ref{conc}. The detailed proofs of theorems and lemmas are presented in the appendix.

\section{Blockchain-based Hierarchical Federated Learning}\label{system}
In this section, we introduce our considered blockchain-based hierarchical federated learning (BHFL) framework, consisting of an HFL system and a blockchain system, where the blockchain is applied to improve efficiency and provide a trustworthy computing environment for the HFL system. Specifically, we discuss the overview of BHFL, the detailed descriptions of the HFL process and blockchain system, and the challenges of stragglers and latency in this framework. 
\subsection{System Overview}

\begin{figure}[htpt]
\centering
\includegraphics[width=0.45\textwidth]{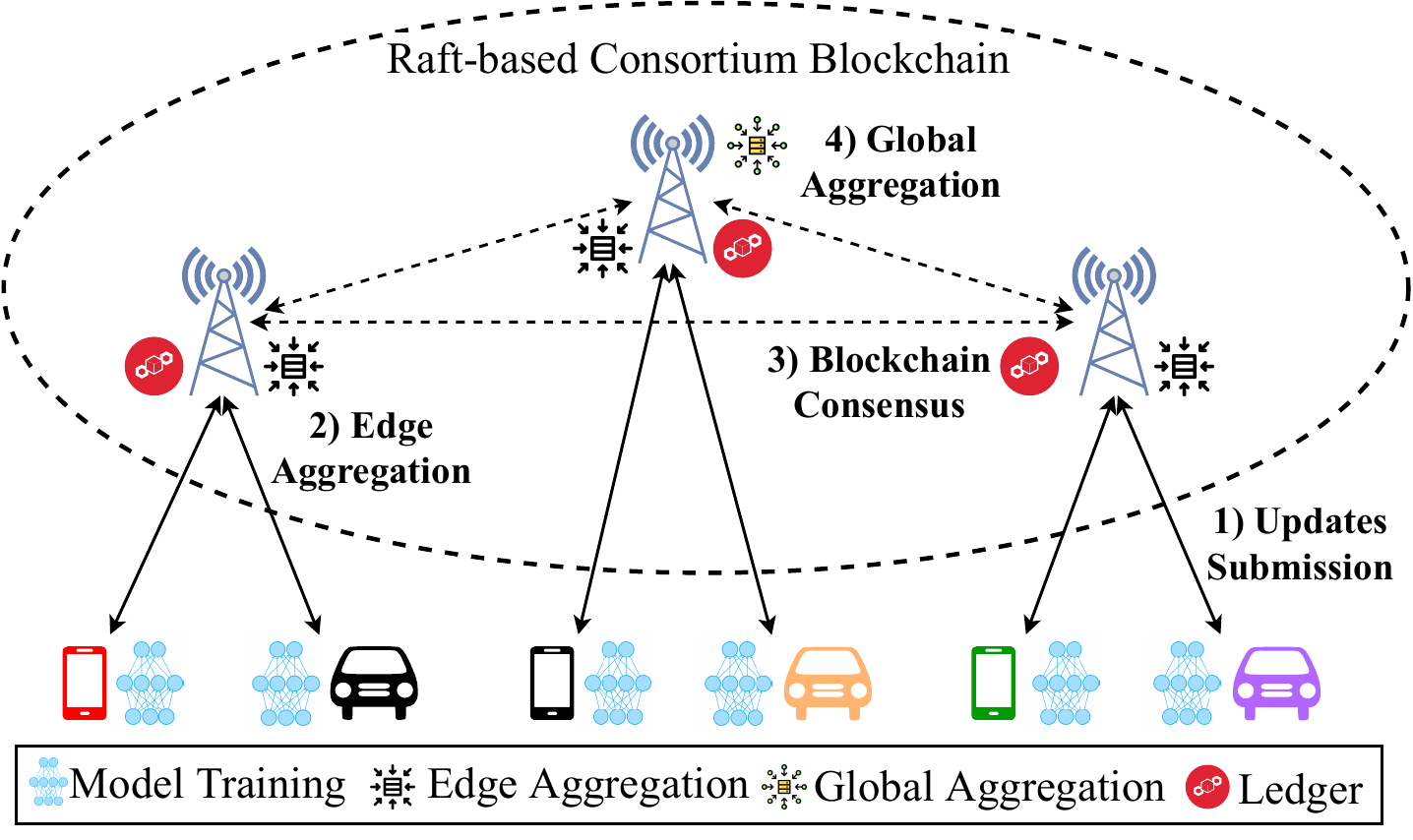}\caption{Blockchain-based Hierarchical Federated Learning.}
\label{fig_topo}
\end{figure}

As shown in Fig. \ref{fig_topo}, the considered BHFL system comprises multiple edge servers and local devices, where a consortium blockchain runs on edge servers. Specifically, local devices and edge servers form several FL systems. 
Let $i\in\left\{1,2,\cdots, N\right\}$ denote the edge server, where $N$ is the total number of edge servers. For edge server $i$, there are $J_i$ local devices connected, and we let $j\in\left\{1,2,\cdots, J_i\right\}$ denote each local device connected with edge server $i$. These devices involved in BHFL are heterogeneous, which means they have various computing and communication resources, as well as different raw data distributions, i.e., non-independent and identically distributed (non-IID) data. 
Denote $k\in\left\{1,2,\cdots, K\right\}$ as the edge aggregation round, i.e., the round of model aggregation on edge servers based on the local model updates from connected devices, where $K$ is the total number of edge aggregation rounds; let $t\in \left\{1,2,\cdots,T\right\}$ be the global aggregation round, i.e., the round of model aggregation on the blockchain system based on the submissions from edge servers, where $T$ is the total number of global aggregation rounds.
We use the pair $(t,k)$ to denote edge aggregation round $k$ in the global aggregation round $t$ and use $(i,j)$ to denote local device $j$ connected to edge server $i$. 

The workflow of our proposed BHFL system can be described below:
\begin{enumerate}
    \item \textbf{Updates Submission}: after multiple local iterations, i.e., the gradients updating of Stochastic Gradient Decent (SGD), local devices submit their trained local models to the connected edge servers.
    \item \textbf{Edge Aggregation}: edge servers aggregate the received local models to get the edge models by our proposed HieAvg which will be detailed in Section \ref{fl}, and return them to local devices for the next round of training till finishing $K$ rounds of edge aggregation.
    \item \textbf{Blockchain Consensus}: while local devices are conducting model training using their own data, edge servers can perform the consensus algorithm in the upper blockchain network, where one edge leader will be elected before global aggregation. 
    (see Section \ref{raft} for details). 
    \item \textbf{Global Aggregation}: after $K$ rounds of edge aggregation, edge servers transmit their latest edge models to the edge leader for global model aggregation by HieAvg, and the edge leader will return the updated global model to edge servers for the next round of training till the BHFL model converges.
\end{enumerate}


\subsection{Hierarchical Federated Learning}\label{system_m}
In FL,  the participants work together to solve a  finite-sum optimization problem with SGD, while in hierarchical FL (HFL), the hierarchical SGD (H-SGD) is adopted\cite{wang2022demystifying}. The main difference between SGD and H-SGD is that H-SGD requires several rounds of intermediate aggregation before global aggregation.

In HFL, we can treat the framework of local devices and their connected edge server $i$ as FL, and its objective function can be expressed as:
\begin{align}
    \quad \mathop{\arg\min}_{w_i^{t,k} \in \mathbb{R}^d}& \quad  F_i(w_i^{t,k})=\frac{1}{J_i}\sum_{j=1}^{J_i} F_{i,j}(w_{i,j}^{t,k})\notag,
\end{align}
where $F_{i,j}(\cdot)$ is the loss function of local device $j$ and $F_{i}(\cdot)$ is the loss function of edge server $i$; and $w_{i,j}^{t,k}$ is the weights of local device $(i,j)$ in round $(t,k)$ and $w_{i}^{t,k}$ is the weights of edge server $i$ in round $(t,k)$.

On local device $(i,j)$, the model is updated by:
\begin{align}
    w_{i,j}^{t,k+1}=w_{i,j}^{t,k}-\eta^{t,k} \nabla F_{i,j} (w_{i,j}^{t,k},\xi_{i,j}^{t,k}),
    \label{e_0}
\end{align}
where $\eta^{t,k}$ is the learning rate in round $(t,k)$; and $\nabla F_{i,j} (w_{i,j}^{t,k},\xi_{i,j}^{t,k})$ is the gradient of $F_{i,j}(w_{i,j}^{t,k})$ with $\xi_{i,j}^{t,k}$ being the random data sample from the raw data of local device $(i,j)$, and we can assume $\mathbb{E}_j[\nabla F_{i,j}(w_{i,j}^{t,k})]=\nabla F_i(w_i^{t,k})$ where $\mathbb{E}(\cdot)$ is the expectation notation.

On the edge leader, the objective function is defined as:
\begin{align}
    \quad \mathop{\arg\min}_{w^t \in \mathbb{R}^d}& \quad  F(w^t)=\frac{1}{N}\sum_{i=1}^{N} F_i(w_i^t)\notag,
\end{align}
where $F(\cdot)$ is the global loss function, $w^t$ is the global weights in round $t$ and $w_i^t$ is the weights of edge server $i$ in round $t$ with $w_i^t=w_i^{t,K}$.

As for the gradient of edge server $i$, we can assume $\mathbb{E}_i[\nabla F_{i}(w_{i}^t)]=\nabla F(w^t)$.
After $T$ rounds of global aggregation, we can get the final global model $w^T$, which should be as approximate as possible to the optimal global model $w^*$ of $F(\cdot)$. 

\subsection{Raft-based Consortium Blockchain}\label{raft}

Generally, there are two ways for synchronizing the model updates in HFL so that the edge servers and local devices can get the latest models in the next round, i.e., \textit{centralized} and \textit{distributed}. In the centralized method, a cloud server is employed to send the latest global model to the edge servers, 
such as the client-edge-cloud framework presented in \cite{liu2020client}; 
while in the distributed way, local model updates are directly shared among peering edge servers via broadcasting, and then each server can derive the global model based on the received updates from others. 
Though the distributed method can eliminate the single point of failure, broadcasting models among edge servers can be costly in communication resource consumption and latency. 

To solve this challenge, blockchain is widely employed to assist in enhancing the efficiency and performance of HFL. In general, two essential requirements have to be satisfied. First, the deployment of blockchain in HFL cannot bring a significant latency increase to the global model convergence, including the latency of computing, communication, and blockchain consensus. Some of the existing studies use computing-intensive blockchain consensus, such as Proof of Work (PoW) \cite{pass2017analysis}, and require model sharing among participants before global model aggregation, making the time cost 
extremely high\cite{wang2021blockchain}. 
Second, the implemented blockchain system needs to be capable of protecting data privacy, i.e., local model updates and global models, from leakage.

For these considerations, we propose a consortium blockchain based on the Raft consensus protocol\cite{ongaro2014search}. Since the consortium blockchain only allows authorized nodes to be included,  the edge servers in this network can thus be trusted; further, as a leader-based consensus mechanism, Raft has been proven to be efficient and reliable. 
The main working process of the Raft-based consortium blockchain in BHFL can be summarized into the following three steps:
\begin{enumerate}
    \item \textbf{Leader Election}: edge servers on the blockchain conduct the leader election process\footnote{For brevity, we omit the detailed leader election process of Raft, which can be found in \cite{ongaro2014search}.}, which should be completed before submitting local models for global aggregation so that the latency of BHFL can be reduced compared to the existing work \cite{wang2021blockchain}. 
    \item \textbf{Model Submission}: edge servers submit their latest edge models to the elected edge leader, and then the leader will aggregate those models to update the global model.
    \item \textbf{Block Generation}: the leader generates a new block that contains all edge models from edge servers and the updated global model, and broadcasts this block to all edge servers on the blockchain. 
\end{enumerate}

Here the Raft-based consortium blockchain in the BHFL system is mainly employed to avoid the single point failure and provide a trustless intermediary for synchronizing model updates among edge servers. 
It is worth noting that the edge servers in the blockchain system are required to finish the leader election process before conducting global aggregation to reduce latency and communication overhead due to broadcasting. As for the details of the latency in blockchain consensus influencing the total system latency, which has been rarely explored in the existing studies, we will discuss them in Section \ref{opt}. 

\subsection{Challenges of BHFL}
As mentioned before, there exist two main challenging factors in BHFL that impact learning performance, i.e., stragglers and latency. We denote $\mathcal{L}_e$ as the predefined waiting period of each edge aggregation round and $\mathcal{L}_g$ as the waiting time of each global aggregation round. The stragglers are BCFL participants that cannot upload models within the required waiting period in both the layers of local devices and edge servers. 
On the one hand, local devices may not submit model updates to edge servers in time; on the other hand, edge servers may miss the required deadline for submitting edge models. These stragglers can result in long latency for the model training. 
In addition to the latency caused by stragglers, the numbers of local devices, edge servers, edge aggregation rounds, and global aggregation rounds, as well as the blockchain consensus process, will influence the total time consumption. 




To resolve the challenge brought by stragglers in BHFL, we first propose a novel model aggregation algorithm, named HieAvg, to mitigate the impact of stragglers in both layers of BHFL, which will be elaborated in Section \ref{fl}. Further, to deal with the challenge of latency, we design an optimization scheme detailed in Section \ref{opt}.

\section{Hierarchical Averaging Aggregation Method }\label{fl}

In this section, we elaborate on our proposed hierarchical averaging (HieAvg) aggregation method. The basic idea of HieAvg is to use the historical weights of stragglers to estimate their delayed weights. 
Generally speaking, there are two main parts of HieAvg: the first part is the basic aggregation method when edge servers would like to collect model submissions from all clients no matter whether there is any straggler or not; 
and the second is the straggler mitigation method with two steps which will be detailed below. 

\subsection{Basic Aggregation Methods of HieAvg}
In the case that edge servers and the edge leader wait for weights from all connected devices without dealing with the stragglers' impact on BHFL convergence, they may use the following aggregation methods to update the edge models and the global model.
\subsubsection{Edge Aggregation}
Let $M_{i}^{t,k}$ be the number of devices that can submit weights to edge server $i$ in round $(t,k)$ within the time requirement, and let $S_i^{t,k}$ denote the number of stragglers among local devices in round $(t,k)$. Thus, we have $J_i=M_i^{t,k}+S_i^{t,k}$, and can get the model of edge server $i$ in round $(t,k)$ as
\begin{align}
    w_{i}^{t,k}=\frac{1}{J_i} \sum_{j=1}^{J_i} w_{i,j}^{t,k}=\frac{1}{J_i} (\sum_{m=1}^{M_i^{t,k}} w_{i,m}^{t,k} + \sum_{s=1}^{S_i^{t,k}} w_{i,s}^{t,k}),
    \label{e_1}
\end{align}
where $w_{i,m}^{t,k}$ and $w_{i,s}^{t,k}$ are the in-time and delayed local weights in round $(t,k)$, respectively.



\subsubsection{Global Aggregation}

We use the following equation to update the global model in round $t$:
\begin{align}
    w^{t}&=\sum_{i=1}^{N} \frac{J_i}{\sum_{i=1}^{N}J_i} w_{i}^{t}=\sum_{m=1}^{M^{t}}\frac{J_m^{t}}{\sum_{i=1}^{N}J_i}w_{m}^{t}+\sum_{s=1}^{S^{t}}\frac{J_s^{t}}{\sum_{i=1}^{N}J_i}w_{s}^{t},
    \label{e_3}
\end{align}
where $M^t$ is the number of edge servers submitting updates to the edge leader timely in round $t$, and $S^t$ is the number of stragglers among edge servers in round $t$;  $J_m^t$ and $J_s^t$ are the numbers of local devices connected to edge server $m$ submitting models in time and that of straggler $s$ in round $t$, respectively; $w_{m}^{t}$ and $w_{s}^{t}$ are the in-time and delayed edge weights in round $t$, respectively.


    
    
    


Note that even if there is no straggler, 
i.e.,  $S_{i}^{t,k}=0$ and $S^{t}=0$, the BHFL system can also apply the above equations to aggregate models.

\noindent \textit{Remark:} Based on the basic aggregation methods of HieAvg, we can find that it differs from FedAvg \cite{mcmahan2017communication} in the following two aspects: i) HieAvg does not require the data size in edge aggregation, avoiding additional data disclosure of local devices; and ii) HieAvg uses the ratio of $J_i$ to the total number of local devices in global aggregation, which is more suitable for HFL, and FedAvg cannot be applied in this situation.


\subsection{Straggler Mitigation of HieAvg}
In this part, we detail the design of HieAvg to mitigate the stragglers' impact, including the steps of \textit{cold boot} and \textit{estimation of delayed weights}.

\subsubsection{Cold Boot}

To better estimate the stragglers' delayed submissions, the BHFL system has to collect enough historical data in the process of cold boot. Denoting $T_c$ as the number of model submision rounds, we require all participants, including local devices and edge servers, to finish at least two rounds of model submission, i.e., $T_c\geq 2$, so that the necessary amount of information can be collected. Ideally, all devices can submit models in time for the first $T_c$ rounds, and the step of cold boot can be described in Algorithm 1. The edge servers need to wait for submissions from local devices for $T_c$ global aggregation rounds (Line 1). 
During cold boot, we use (\ref{e_1}) and (\ref{e_3}) to aggregate the models on edge servers and the edge leader, respectively (Lines 2-13).

\begin{algorithm}
\caption{Cold Boot in HieAvg} 
\label{al_1}
\begin{algorithmic}[1]
\REQUIRE $T_c$, $N$, $K$, $J_i$, $\eta^{t,k}$
\ENSURE $w^{T_c}$

\FOR{$t\in \{1,\cdots,T_c\}$}
\FOR{$i\in \{1,\cdots,N\} ~\mathrm{parallelly} $}
\FOR{$k \in \{1,\cdots,K\}$}
\FOR{$j \in \{1,\cdots,J_i\}~ \mathrm{parallelly}$}
\STATE $w_{i,j}^{t,k}$ $\leftarrow$ updated by  (\ref{e_0})
\ENDFOR
\STATE $w_{i}^{t,k}$ $\leftarrow$ updated by  (\ref{e_1})
\ENDFOR
\STATE $w_{i}^{t}\leftarrow w_{i}^{t,K}$
\ENDFOR
\STATE $w^{t}$ $\leftarrow$ updated by  (\ref{e_3})
\ENDFOR
\RETURN $w^{T_c}$
\end{algorithmic}
\label{al_11}
\end{algorithm}

In the case that one device loses connection after the first round of model submission while other devices can continue working, if the device is reconnected and submits weights after multiple rounds, the resubmitted weights will be considered as the historical weights.

\if()
In this code boot step, we exclude devices that fail to submit models for a long time due to network connectivity problems, assuming that their networks are unstable, thus reducing the risk of causing excessive latency during the training process. In addition, we can choose a smaller value of $T_c$ to reduce the latency of cold boot.
\fi

\subsubsection{Estimation of Delayed Weights}\label{cdw}
After cold boot, we design the scheme of delayed weight estimation to mitigate the impact of stragglers by estimating their delayed weights via the historical weights.









\textbf{Estimation of Delayed Local Weights:}\label{cc1}
Since the edge servers have the historical weights of stragglers, we can use those weights to estimate the delayed weights of stragglers.
We have to ensure that the difference between the estimated weights and the real delayed weights is as small as possible. To that aim, we design an approximate method by utilizing the difference in historical weights of stragglers to estimate their delayed weights in round $(t,k)$. The estimation of delayed weights is:
\begin{align}
    \overline{w}_{i,s}^{t,k}=w_{i,s}^{t,k-1}+ \mathbb{E}_k[\Delta_{i,s}^{t,k-1}],\notag
\end{align}
where
$\Delta_{i,s}^{t,k-1}=w_{i,s}^{t,k-1}-w_{i,s}^{t,k-2}$, and $\mathbb{E}_k[\Delta_{i,s}^{t,k-1}]$ is the expectation of $\Delta_{i,s}^{t,k-1}$ used to avoid large estimation bias.

Then, the estimated $w_{i}^{t,k}$ can be written as:
\begin{align}
    \overline{w}_i^{t,k}&=\frac{1}{J_i} \bigg[\sum_{m=1}^{M_i^{t,k}} w_{i,m}^{t,k} +  \sum_{s=1}^{S_i^{t,k}}\gamma_{i,s}^{t,k}(w_{i,s}^{t,k-1}+\mathbb{E}_k[\Delta_{i,s}^{t,k-1}])\bigg],
    \label{e_2}
\end{align}
where $\gamma_{i,s}^{t,k} =\gamma_0\lambda^{k'}$ is the decay factor used to scale estimated delayed weights with $\gamma_0 \in (0,1)$ being the initial decay factor, $\lambda \in (0,1)$ being the scalar, and $k'\geq 1$ being the missing edge aggregation rounds of stragglers. 









\textbf{Estimation of Delayed Edge Weights:}\label{cc2}
As for stragglers among edge servers, we use the same estimation method for dealing with stragglers among local devices. 
Then, we can get the estimated $w^t$ by the following equation:
\begin{align}
\overline{w}^t&=
    \sum_{m=1}^{M^{t}}\frac{J_m^{t}}{\sum_{i=1}^{N}J_i}w_m^{t}+\sum_{s=1}^{S^{t}}\frac{\gamma_s^t J_s^{t}}{\sum_{i=1}^{N}J_i}(w_s^{t-1}+\mathbb{E}_t[\Delta_s^{t-1}]),
    \label{e_4}
\end{align}
where $\gamma_s^t=\gamma_0\lambda^{t'}$ is the delay factor with $t'$ being the missing global aggregation rounds of stragglers; and $\Delta_s^{t-1}=w_s^{t-1}-w_s^{t-2}$.


The process of estimating delayed weights in HieAvg is detailed in Algorithm \ref{al_1}. If there are stragglers, the corresponding edge server and the edge leader will use the estimation method to update the models (Line 4 and Line 9). Please note that this algorithm is used to handle situations where there exist stragglers; while if there are no stragglers, the model updating will be the same as Algorithm 1. Here we can see that the time complexity of HieAvg, including Algorithms 1 and 2,  is $\mathcal{O}(T\times N\times K \times J)$.

\begin{algorithm}
\caption{Estimation of Delayed Weights in HieAvg} 
\label{al_1}
\begin{algorithmic}[1]
\REQUIRE $T$, $T_c$, $N$, $K$, $J_i$, $\gamma_0$, $\lambda$, $\eta^{t,k}$
\ENSURE $w^T$

\FOR{$t\in \{T_c+1,\cdots,T\}$}
\FOR{$i\in \{1,\cdots,N\} ~\mathrm{parallelly} $}
\FOR{$k \in \{1,\cdots,K\}$}
\STATE $\overline{w}_{i}^{t,k}$ $\leftarrow$ updated by  (\ref{e_2})
\STATE $w_{i}^{t,k}$ $\leftarrow$ $\overline{w}_{i}^{t,k}$
\ENDFOR
\STATE $w_{i}^{t}$ $\leftarrow$ $w_{i}^{t,K}$
\ENDFOR
\STATE $\overline{w}^{t}$ $\leftarrow$ updated by  (\ref{e_4})
\STATE $w^{t}$ $\leftarrow$ $\overline{w}^{t}$
\ENDFOR
\RETURN $w^T$
\end{algorithmic}
\label{al_11}
\end{algorithm}


In fact, there exist two types of stragglers for both the stragglers among local devices and edge servers in BHFL: \textit{permanent stragglers} and \textit{temporary stragglers}. Without loss of generality, we take the stragglers among edge servers as an example to clarify this point. First, if the stragglers will never return to join the BHFL training process due to the loss of communication connection or location change at global round $t$, then we can use the above method to estimate the updates of stragglers starting from round $t$ to the end of training. We call this kind of stragglers permanent stragglers. Second, if the stragglers will return after $t'\geq 1$ rounds, we can still first use the above method to estimate the updates of stragglers during rounds $t$ to $t+t'$, and once the stragglers return in round $t+t'+1$, they can submit their latest models. These stragglers are named temporary stragglers. Intuitively, permanent stragglers are more harmful to BHFL than temporary stragglers since the bias will be larger if the stragglers disappear. Thus, we treat the permanent stragglers as the worst case in the following section of convergence analysis.






\section{Convergence Analysis of HieAvg-based BHFL }\label{convg}
In this section, we analyze the convergence of BHFL with HieAvg. We first introduce the necessary assumptions for theoretical proof and then discuss the convergence of edge aggregation and global aggregation subsequently.

\subsection{Assumptions}
Here we introduce two  assumptions that are important for the proof of convergence. The first one indicates the property of the loss function employed in our proposed BHFL framework, which has also been widely included in the existing studies \cite{wang2022demystifying, liu2022accelerating, nguyen2022latency}. The second ensures that the model updating process will not lead to a significant bias.


\begin{assum}
(Lipschitz-smoothness)
The loss function $F(\cdot)$ is continuously differentiable  and the gradient function of $F(\cdot)$ is Lipschitz continuous with Lipschitz constant $L>0$, which means $|| \nabla F(w)-\nabla F(\overline{w})||^2\leq L || w-\overline{w}||^2$ for all $w, \overline{w}\in \mathbb{R}$. It also implies that
\begin{align}
    F(w)-F(\overline{w})\leq \nabla F(w)^T (\overline{w}-w)+\frac{L}{2} ||w-\overline{w}||^2,\notag
\end{align}
where $||\cdot||^2$ is the $l_2$ norm.
\end{assum}

\begin{assum}(Bounded Variance) Three types of bounded variance are assumed: 

1) Bounded Variance of Weight Difference: 
\begin{gather}
    \mathbb{E}_k[ || (w_{i,j}^{t,k}-w_{i,j}^{t,k-1})-\Delta_{i,j}||^2]\leq \delta_{i,j}^2,\notag\\
    \mathbb{E}_t[ || (w_{i}^t-w_{i}^{t-1})-\Delta_{i}||^2]\leq \delta_{i}^2,\notag
\end{gather}
where $\Delta_{i,j}=\mathbb{E}_k[\Delta_{i,j}^{t,k-1}]$ and $\Delta_{i}=\mathbb{E}_t[\Delta_i^{t}]$; and $\delta_{i,j}, \delta_{i} \in \mathbb{R}^+$.

2) Bounded Variance of Estimated Gradients:
\begin{gather}
    \mathbb{E}_j[||\nabla F_{i,j}(\overline{w}_{i,j}^{t,k})-\nabla F_i(\overline{w}_i^{t,k})||^2] \leq \delta'^2,\notag\\
    \mathbb{E}_i[||\nabla F_{i}(\overline{w}_{i}^{t})-\nabla F(\overline{w}^{t})||^2] \leq \delta''^2,\notag
\end{gather}
where $\nabla F_i(\overline{w}_i^{t,k})=\mathbb{E}_j[\nabla F_{i,j}{(\overline {w}_{i,j}^{t,k})}]$ and $\nabla F(\overline{w}^t)=\mathbb{E}_i[\nabla F_{i}(\overline{w}_{i}^t)]$; and $\delta', \delta'' \in \mathbb{R}^+$.

3) Bounded Variance of Estimated Delayed Weights:
\begin{gather}
    \mathbb{E}_i[ || \overline{w}_{i}^{t,k}-\overline{w}^{t,k}||^2]\leq \overline{\delta}^2,\notag\\
    \mathbb{E}_t[ || \overline{w}^t-\overline{w}||^2]\leq \overline{\delta}'^2,\notag
\end{gather}
where $\overline{w}^{t,k}=\mathbb{E}_i[\overline{w}_{i}^{t,k}]$ is the auxiliary variable inspired by \cite{stich2018local,li2019convergence}; and $\overline{w}=\mathbb{E}_t[\overline{w}^t]$; and
$\overline{\delta}, \overline{\delta}' \in \mathbb{R}^+$.
\end{assum}


Assumption 2.1 is unique in this work since we use the difference of weights to estimate the delayed weights, which is assumed to have bounded variance; and Assumptions 2.2 and 2.3 are about the estimated weights, guaranteeing that the estimation method will not lead to significant bias. 

It is worth noting that the learning rate $\eta^{t,k}=\frac{1}{\eta_0+d(tK+k)}$ is assumed to be dynamic, where $\eta_0$ is the initial learning rate for all the local devices and $d$ is the decay rate. Besides, we have no assumption on the convexity of the loss function; however, since the non-convex case is more challenging, we will analyze the convergence of HieAvg in BHFL with the non-convex loss function in the below. 


%
\subsection{Convergence of HieAvg}
Before we discuss the convergence of HieAvg on both layers, we need to introduce two useful lemmas which will be applied in the proof of convergence.
\begin{lemma}
Under Assumption 2, by applying HieAvg on edge servers, the difference between the estimated edge model in round $(t,k+1)$ and that in round $(t,k)$ is bounded by
\begin{align}
    &\overline{w}_{i}^{t,k+1}-\overline{w}_i^{t,k}\leq \overline{\delta}-\eta^{t,k} \nabla F_i(\overline{w}_i^{t,k})-\gamma_0\frac{S_i^{t,k}}{J_i} (\Delta_{i,j}+ \delta_{i,j}^2),\notag
\end{align}
where 
$\frac{S_i^{t,k}}{J_i}$ denotes the proportion of stragglers among local devices connected to edge server $i$ in round $(t,k)$.
\end{lemma}

\begin{lemma}
Under Assumption 2, by applying HieAvg on the edge leader, the difference between the estimated global model in round $t+1$ and that in round $t$ is bounded by
\begin{align}
    \overline{w}^{t+1}-\overline{w}^{t}&\leq \overline{\delta}'-[\frac{\mathbb{E}_s[J_s^t]}{N\mathbb{E}_i[J_i]} + \gamma_0 \frac{S^t}{N}( \Delta_{i}+ \delta_{i}^2)]\notag\\
    &-\frac{K \mathbb{E}_s[J_s^t]}{N\mathbb{E}_i[J_i]} \eta^{t,k} \nabla F(\overline{w}^{t}),\notag
\end{align}
where $\frac{S^t}{N}$ is the proportion of stragglers among edge servers in round $t$.
\end{lemma}

Both Lemma 1 and Lemma 2 imply that the difference in estimated weights will be affected by the previous weight differences, the delayed weights, and the proportion of stragglers. By now, however, it is still unclear how these factors influence the convergence of HieAvg based on the above two lemmas, which will be explored in the following two subsections. 

\subsubsection{Convergence on Edge Servers}
We first investigate the convergence of HieAvg on edge servers. By analyzing the convergence on edge servers, we can see the effectiveness of our proposed HieAvg algorithm.
\begin{theorem}
Under Assumption 1 and Assumption 2, with dynamic learning rate $\eta^{t,k}$, the number of stragglers $S_i^{t,k}$, and the number of connected local devices $J_i$, if $\eta^{t,k}>\frac{1}{L+2}$ with $L> 0$ and $\gamma_0 \frac{\mathbb{E}_k[S_i^{t,k}]}{J_i} (\Delta_{i,j}+ \delta_{i,j})-\overline{\delta}\geq0$, by applying Lemma 1, the convergence of HieAvg on edge server $i$ is bounded by  
\begin{align}
   & \frac{1}{K}\sum_{k=1}^{K}\mathbb{E}[|| \nabla F_i (\overline{w}_i^{t,k})||^2]\notag\\
   \leq & \frac{2[F_i(w_i^0)-F_i(w_i^*)+\frac{2\mathbb{E}_k[\eta^{t,k}]\delta'^2}{L\mathbb{E}_k[\eta^{t,k}] +2\mathbb{E}_k[\eta^{t,k}]-1}]}{(L\mathbb{E}_k[\eta^{t,k}] +2\mathbb{E}_k[\eta^{t,k}]-1) \sqrt{K}}\notag\\
   &+\frac{(2+L) [\gamma_0 \frac{\mathbb{E}_k[S_i^{t,k}]}{J_i} (\Delta_{i,j}+ \delta_{i,j})-\overline{\delta}] }{L\mathbb{E}_k[\eta^{t,k}] +2\mathbb{E}_k[\eta^{t,k}]-1},\notag
\end{align}
where $w_i^0$ is the initial weights of edge server $i$ and $w_i^*$ is the optimal weights of edge server $i$.
\end{theorem}

The above inequality provides a theoretical upper bound for the averaging expectation of squared gradient norms of $F_i(\cdot)$, which indicates that the loss function of edge $i$ can converge to a critical point with enough edge aggregation rounds, smaller learning rate, and fewer stragglers among local devices.
Besides, Theorem 1 can be employed to analyze the convergence of traditional FL, 
which is guaranteed to converge even with the existence of stragglers if $K$ is well selected.


\subsubsection{Convergence on the Edge Leader}
Now, we can discuss the global convergence on the blockchain. 
\begin{theorem}
Under Assumption 1 and Assumption 2, with dynamic learning rate $\eta^{t,k}$, the number of stragglers $S^{t}$, and the number of connected local devices $J_i$, if $\eta^{t,k}\geq\frac{1}{L+\frac{2 K \mathbb{E}_t[J_s^t]}{ N\mathbb{E}_i[J_i]}}$ with $L> 0$ and $\frac{\mathbb{E}_t[J_s^t]}{N\mathbb{E}_i[J_i]}+ \gamma_0 \frac{\mathbb{E}_t[S^t]}{N}( \Delta_{i}+ \delta_{i}^2)-\overline{\delta}'\geq 0$, by applying Lemma 2,  the convergence of HieAvg on the edge leader is bounded by
\begin{align}
    &\frac{1}{T}\sum_{t=1}^{T}\mathbb{E}[||\nabla F(\overline{w}^t)||^2]\notag\\
    &\leq\frac{2 [F(w^{0})-F(w^*)+\frac{\sqrt{K} \mathbb{E}_t[\eta^{t,k}] \mathbb{E}_t[J_s^t]}{N\mathbb{E}_i[J_i]}\delta''^2]}{\sqrt{T}(2\sqrt{K}\frac{ \mathbb{E}_t[\eta^{t,k}] \mathbb{E}_t[J_s^t]}{N\mathbb{E}_i[J_i]} +L\mathbb{E}_t[\eta^{t,k}] -1)}\notag\\
    &+\frac{(2+L) [\frac{\mathbb{E}_t[J_s^t]}{N\mathbb{E}_i[J_i]}+ \gamma_0 \frac{\mathbb{E}_t[S^t]}{N}( \Delta_{i}+ \delta_{i}^2)-\overline{\delta}']}{2 \sqrt{K}\frac{ \mathbb{E}_t[\eta^{t,k}] \mathbb{E}_t[J_s^t]}{N\mathbb{E}_i[J_i]} +L\mathbb{E}_t[\eta^{t,k}] -1},\notag
\end{align}
where $w^0$ is the initial global weights; for convenience, we use $\Omega$ to represent the upper bound at the right side of the above inequality.
\end{theorem}

Based on the above theorem, we can see that the averaging expectation of squared gradient norms of $F(\cdot)$ has an upper bound, which implies that BHFL with HieAvg can converge.
Besides, by analyzing Theorem 1 and Theorem 2, we can obtain the following two corollaries to further explain the convergence performance of HieAvg.


\begin{corollary}\label{c3}
Given the fixed values of other influence factors, the convergence performance can be better achieved with more edge aggregation rounds ($K$). 
\end{corollary}
Corollary 1 indicates that we can speed up global convergence with more edge aggregation rounds. This is because if there are more rounds of edge aggregation, each edge server can get a model with a smaller loss,  which accelerates the convergence of the global model during the phase of global aggregation.

\begin{corollary}\label{c2}
Given the fixed values of other influence factors, the convergence performance can be better achieved with fewer stragglers in local devices and edge servers ($S_i^{t,k}$ and $S^t$).
\end{corollary}


Corollary 2 demonstrates the influence of stragglers on the convergence of HieAvg. The occurrence of stragglers is usually caused by unpredictable network conditions, and it is nearly impossible to eliminate their effects on model training completely; but with our proposed HieAvg, the convergence of BHFL is guaranteed.

In summary, the HieAvg algorithm is convergence-guaranteed with a non-convex loss function and non-IID data 
even when there are stragglers among local devices and edge servers in BHFL.





\section{Latency Optimization of BHFL}\label{opt}
In this section, we target to resolve the challenge of latency in BHFL by studying the latency optimization of our proposed framework.
\subsection{Latency Model}\label{subsec:latency model}
\subsubsection{Latency of Local Devices}
By applying Shannon's theory\cite{shannon1948mathematical}, we can calculate the data transmission rate of local device $j$ connected to edge server $i$ in round $(t,k)$ as $r_{i,j}^{t,k}=B_{i,j}^{t,k}\log_2 (1+\frac{u_{i,j}^{t,k} \pi_{i,j}^{t,k}}{\epsilon^2})$,  
where $B_{i,j}^{t,k}$ is the bandwidth of local device $(i,j)$ in round $(t,k)$; $u_{i,j}^{t,k}$ and $\pi_{i,j}^{t,k}$ are the transmission power and channel power gain, respectively; and $\epsilon$ is the Gaussian noise. Then, the transmission time of one communication round between the local device $j$ and edge server $i$ can be calculated by $\mathcal{LM}_{i,j}^{t,k}=\frac{D_{i,j}^{t,k}}{r_{i,j}^{t,k}}$, 
where $D_{i,j}^{t,k}$ is the size of local model updates.

The computing latency before each round of edge aggregation can be computed as $\mathcal{LP}_{i,j}^{t,k}=\frac{C_{i,j}^{t,k}}{f_{i,j}^{t,k}}$,
where $C_{i,j}^{t,k}$ is the total CPU cycles required to complete the training in edge round $(t,k)$  and $f_{i,j}^{t,k}$ is the unit CPU cycles of local device $(i,j)$. 

Since the communication between the local device $j$ and edge server $i$ includes both model downloading and model update submission, the total latency\footnote{This latency can also utilize the value of the slowest device, and the main solution proposed in this section can be applied similarly.} on local devices during one round of edge aggregation is 
\begin{align}
    \mathcal{L}_{lc}=\sum_{t=1}^T\sum_{i=1}^{N} \sum_{k=1}^K\sum_{j=1}^{J_i}(2\mathcal{LM}_{i,j}^{t,k} + \mathcal{LP}_{i,j}^{t,k}).\notag
\end{align}


\subsubsection{Latency of Edge Servers} On edge servers, they are mainly responsible for intermediate model aggregation and transmission. Here we omit the time consumption of model aggregation since it is negligible compared to that of model transmission.
Similarly to the calculation of the communication latency of local devices, we can get the total communication latency of servers as 
\begin{align}
    \mathcal{L}_{gb}=2 \sum_{t=1}^T\sum_{i=1}^{N} \mathcal{LM}_{i}^{t},\notag
\end{align}
 where $\mathcal{LM}_{i}^{t}$ is the communication time cost for model uploading and downloading of edge server $i$. 
\subsubsection{Latency of Blockchain Consensus} Let $\mathcal{L}_{bc}$ be the latency of blockchain consensus in each global round, and denote 
\begin{align}
    \mathcal{L}_{g}=K \max_{t\leq T_c}(\mathcal{LM}_{i,j}^{t,k}+\mathcal{LP}_{i,j}^{t,k}),\notag
\end{align}
 as the waiting period for round $t$. Then $\mathcal{L}_{bc}\leq \mathcal{L}_{g}$ becomes a constraint for the Raft-based blockchain system to guarantee that its deployment brings no increase to the overall latency of BHFL.

\subsubsection{Total Latency} The total latency, denoted as $\mathcal{L}$, is the sum of the latency of local devices and edge servers. Note that the latency of blockchain consensus is not included in the total latency because the blockchain consensus has been completed during $K$ rounds of edge aggregation as required above. 
Thus, we have:
\begin{align}
    \mathcal{L}=\mathcal{L}_{lc}+\mathcal{L}_{gb}&=\sum_{t=1}^T\sum_{i=1}^{N} \sum_{k=1}^K\sum_{j=1}^{J_i}(2\mathcal{LM}_{i,j}^{t,k} + \mathcal{LP}_{i,j}^{t,k})\notag\\
    &+2 \sum_{t=1}^T\sum_{i=1}^{N}\mathcal{LM}_{i}^{t}.\notag
\end{align}


For a rough qualitative analysis, we assume that the number of local devices connected to each edge server is the same, which is denoted by $J$; and we can assume the latency of each local device is fixed in each round, and thus we use $\mathcal{LM}_{i,j}$ to represent the communication latency of local device $(i,j)$; similarly, we use $\mathcal{LP}_{i,j}$ and $\mathcal{LM}_{i}$ to stand for the computing latency of local device $(i,j)$ and the latency of edge server $i$, respectively.
Then, we can simplify the above equation as 
\begin{align}
    \mathcal{L}\approx TNJK(2\mathbb{E}[\mathcal{LM}] + \mathbb{E}[\mathcal{LP}])+2TN\mathbb{E}[\mathcal{LM}'],\notag
\end{align}
 where $\mathbb{E}[\mathcal{LM}]=\mathbb{E}_i[\mathbb{E}_j[\mathcal{LM}_{i,j}^{t,k}]]$, $\mathbb{E}[\mathcal{LP}]=\mathbb{E}_i[\mathbb{E}_j[\mathcal{LP}_{i,j}^{t,k}]]$ and $\mathbb{E}[\mathcal{LM}']=\mathbb{E}_i[\mathcal{LM}_{i}^t]$.  We can see that $\mathcal{L}$ 
and $K$ are positively proportional, and thus we can conclude that reducing the frequencies of edge aggregation can lead to lower communication latency. However, we know that larger $K$ will improve the convergence performance of BHFL according to Theorem 2. Thus, $K$ should be determined by jointly considering the performance and latency of BHFL.
\subsection{Latency Optimization}\label{opt_o}
In this part, we formulate the latency optimization problem by reducing latency and maintaining the convergence performance of BHFL at the same time. Based on the above analysis, we know that $\mathcal{L}$ is a linear function of $K$ if we use the expectations  of communication and computing latency to calculate $\mathcal{L}$, so we can approximately get the optimal $K$ by solving the following optimization problem:
\begin{align}
   \quad \mathop{\arg\min}_{K}& \quad  \mathcal{L}\notag\\
   \text{s.t.}
~&\mathrm{C1}:\Omega \leq \overline{\Omega},\notag\\
& \mathrm{C2}: \mathcal{L}_{bc}\leq \mathcal{L}_{g},\notag\\
& \mathrm{C3}:K\in \mathbb{N^+}\notag,
\end{align}
where C1 is the constraint of convergence performance, ensuring that BHFL can have a good performance to meet the requirement $\overline{\Omega}$; and C2 constrains the waiting time in each global round by considering the time consumption of blockchain consensus; and C3 indicates that $K$ should be a positive integer. Then the above optimization problem becomes a simple integer linear programming with inequality constraints, which can be resolved using classical solutions with polynomial complexity, such as CVXPY\cite{diamond2016cvxpy}, 
to find the optimal number of edge aggregation rounds, i.e., $K^*$.

\section{Experimental Evaluation}\label{exp}
In this section, we evaluate our proposed HieAvg algorithm and latency optimization scheme via extensive experiments. We first introduce the experimental settings and then present the experimental results with discussions. 
\subsection{Experimental Settings}
\subsubsection{BHFL Basic Setting}
Unless specified otherwise, we use the following basic setting in our experiments. 
We simulate a BHFL framework with five edge servers, where each edge server is connected to five local devices. There are two edge aggregation rounds between two rounds of global aggregation, i.e., $K=2$. Each local device owns at most one class of data. We assume there are 20\% stragglers in each layer, which means that one edge server cannot submit the edge model timely in each round of global aggregation  
and one local device connected to each edge server fails to upload the local model timely in each edge aggregation round, respectively. Besides, we set $\gamma_0=0.9$ and $\lambda=0.9$ for HieAvg.
\subsubsection{Stragglers}
For permanent stragglers, they stop submitting model updates after 40 rounds.
And temporary stragglers miss submissions in multiple single rounds but will continue to submit in the next round after the missing round.

\subsubsection{Dataset}
We use MNIST \cite{mnist} as the example dataset in BHFL, which contains 70,000 handwritten digits from 0 to 9. When there are no stragglers, the accuracy is about 87.75\%. 

\subsubsection{Machines and Platforms}
We develop our proposed BHFL framework based on Python 3.7 and TensorFlow 2.9 on Google Colab, Raspberry Pi 4 Model B, and AWS EC2. Specifically, we test the convergence of BHFL on Colab with an A100 GPU and explore the latency of communication for model synchronization with Raspberry Pi and AWS EC2.

\subsubsection{Learning Models}
Based on TensorFlow, we create a CNN-based deep learning model with two convolutional layers, one max pooling layer, one flattening layer, and one dense layer. The batch size is 32, the local iteration is one epoch, and the initial learning rate is 0.001 with the decay rate $d=0.90$.

\subsubsection{Benchmarks of Aggregation Methods}
We consider three benchmarks based on federated averaging (FedAvg)\cite{mcmahan2017communication} to compare with our proposed BHFL framework with HieAvg from the convergence perspective.
The first benchmark considers no stragglers, which is named as \textit{W/O Stragglers}. For the second solution dealing with stragglers, only the timely submissions from local devices and edge servers will be included in edge aggregation and global aggregation, which is termed as \textit{T\_FedAvg}. 
The third one uses the weights submitted in the last round as the weights of stragglers in round $k$ or $t$, which is called \textit{D\_FedAvg}.

\subsubsection{Computing and Communication}
We calculate the computing and communication latency on machines and platforms according to the equations in Section \ref{subsec:latency model}. Specifically, we simulate the model training process of one local device on Raspberry Pi, and we let the Raspberry Pi communicate with EC2 to get the latency of communication.

\subsection{Experimental Results}
\subsubsection{Evaluation of Convergence}

\begin{figure}[h]
\centering
\subfigure[Permanent Stragglers.]{
\includegraphics[width=0.23\textwidth]{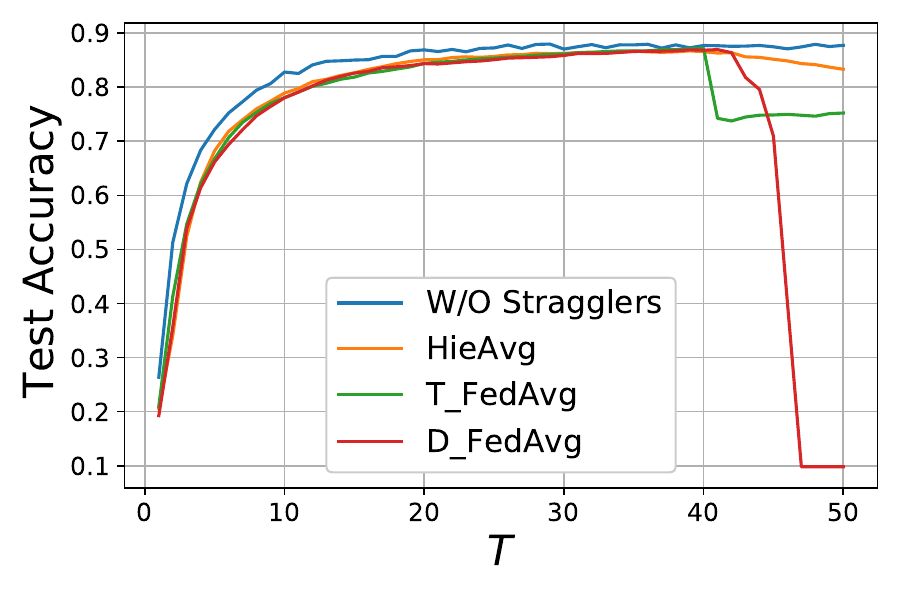}}
\subfigure[Temporary Stragglers.]{
\includegraphics[width=0.23\textwidth]{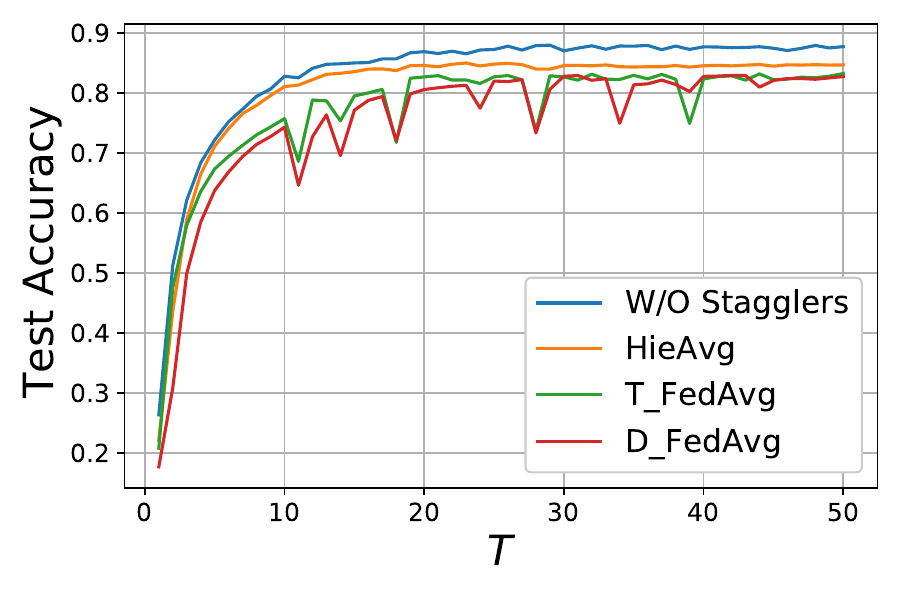}}
\caption{Comparison with Different Aggregation Algorithms.
}
\label{fig_e}
\end{figure}

We first compare the performance of different algorithms in handling both permanent and temporary stragglers. 
The results are shown in Fig. \ref{fig_e}. From Fig. \ref{fig_e}(a) involving permanent stragglers, we can see that compared to the ideal case, i.e., W/O Stragglers, the other three algorithms have various losses of accuracy. 
The accuracy of T\_FedAvg decreases a lot, and D\_FedAvg fails to converge, while our proposed HieAvg can still have relatively good accuracy in handling permanent stragglers. In Fig. \ref{fig_e}(b) dealing with temporary stragglers, all algorithms can achieve good accuracy, but the convergence of HieAvg is smoother and faster than T\_FedAvg and D\_FedAvg. These two sets of experiments illustrate that different kinds of stragglers affect global convergence, but the proposed HieAvg performs better in both cases.

\begin{figure}[h]
\centering
\subfigure[Accuracy with Different $J$.]{
\includegraphics[width=0.23\textwidth]{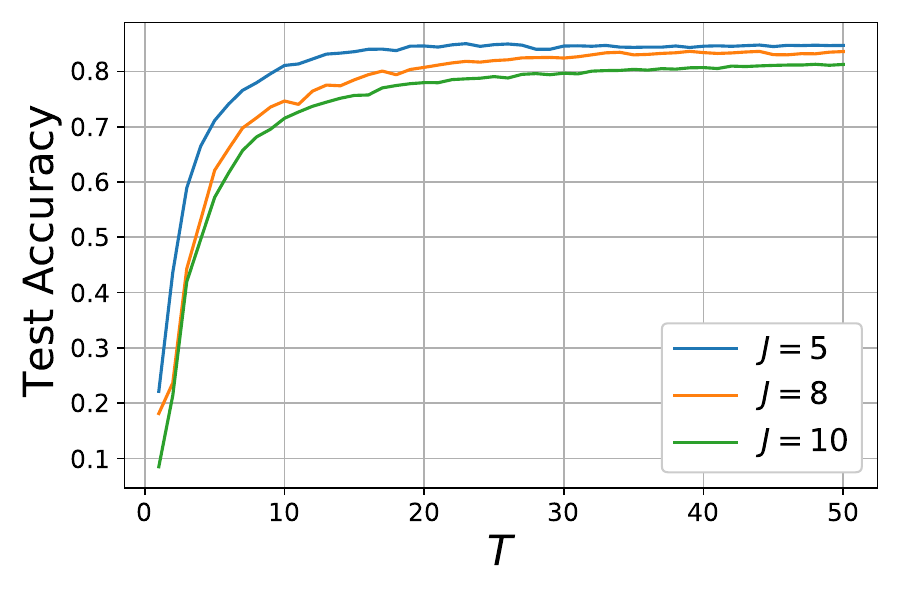}}
\subfigure[Accuracy with Different $N$.]{
\includegraphics[width=0.23\textwidth]{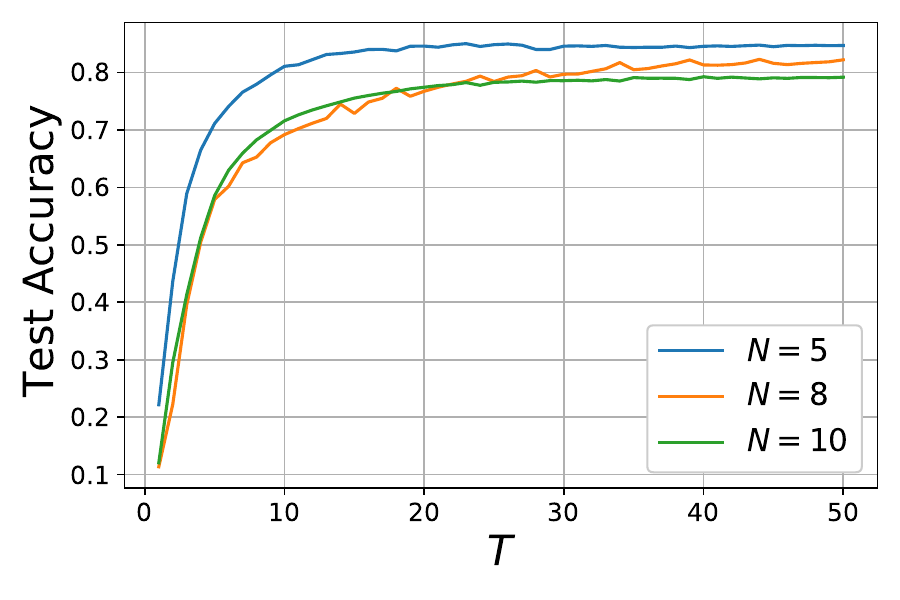}}
\subfigure[Accuracy with Different $K$.]{
\includegraphics[width=0.23\textwidth]{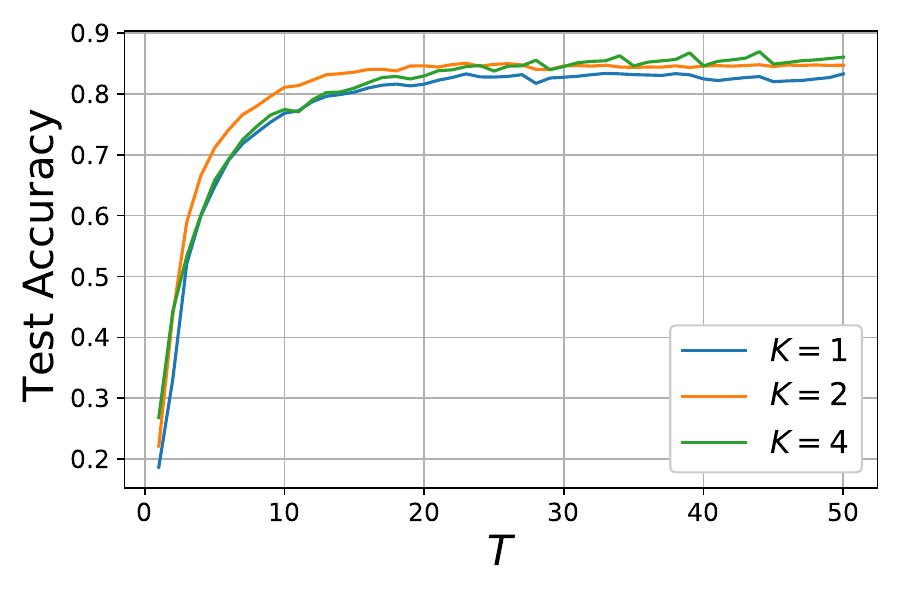}}
\subfigure[Accuracy with Different $S_i$ and $S_{i,j}$.]{
\includegraphics[width=0.23\textwidth]{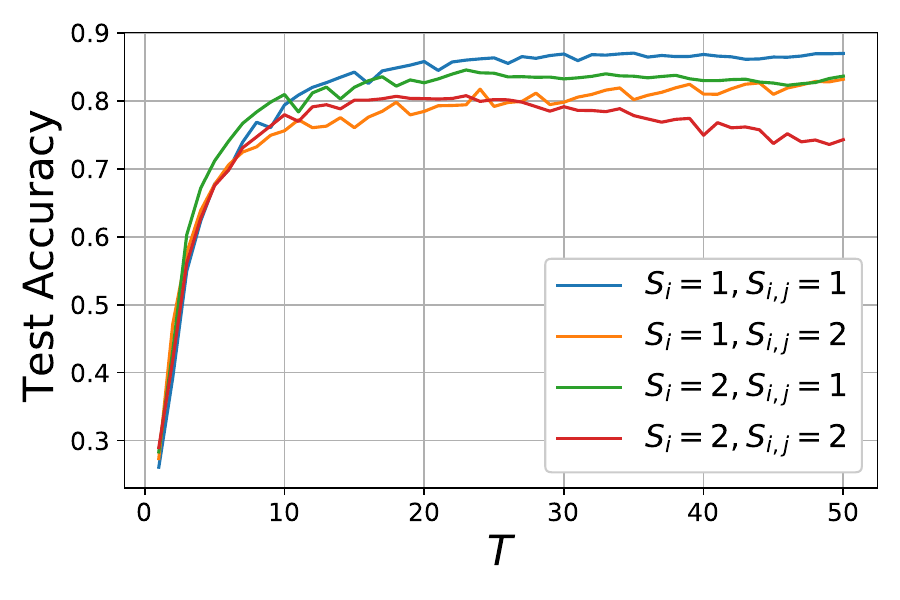}}
\caption{Influences of Parameters on BHFL Training.}
\label{fig_h}
\end{figure}
Next, we explore the impact of different parameter settings, including the numbers of local devices, edge servers, edge aggregation rounds, and stragglers in two layers, on HieAvg with temporary stragglers. By changing $J$, we obtain the results shown in Fig. \ref{fig_h}(a), which indicates that HieAvg converges faster when there are fewer local devices. By adjusting the number of edge servers, as shown in Fig. \ref{fig_h}(b), we can get a similar conclusion. This is because increasing the number of local devices and edge servers will aggravate the imbalance of data distribution when the total data volume of MNIST is fixed under the non-IID situation, thus leading to performance degradation. Later, we analyze the influence of $K$ on the accuracy. Fig. \ref{fig_h}(c) implies that more edge aggregation rounds help improve the accuracy because the more frequent edge aggregation allows each edge server to better integrate the data characteristics of local devices for local optimization. With the varying number of stragglers, the results are reported in Fig. \ref{fig_h}(d), showing that as the number of stragglers increases, the model performance decreases. However, even in the case of 40\% being stragglers (i.e., $S_i=2, S_{i,j}=2$), HieAvg can still achieve an accuracy of 0.74. 

\begin{figure}[h]
\centering
\subfigure[Different Data Distributions.]{
\includegraphics[width=0.23\textwidth]{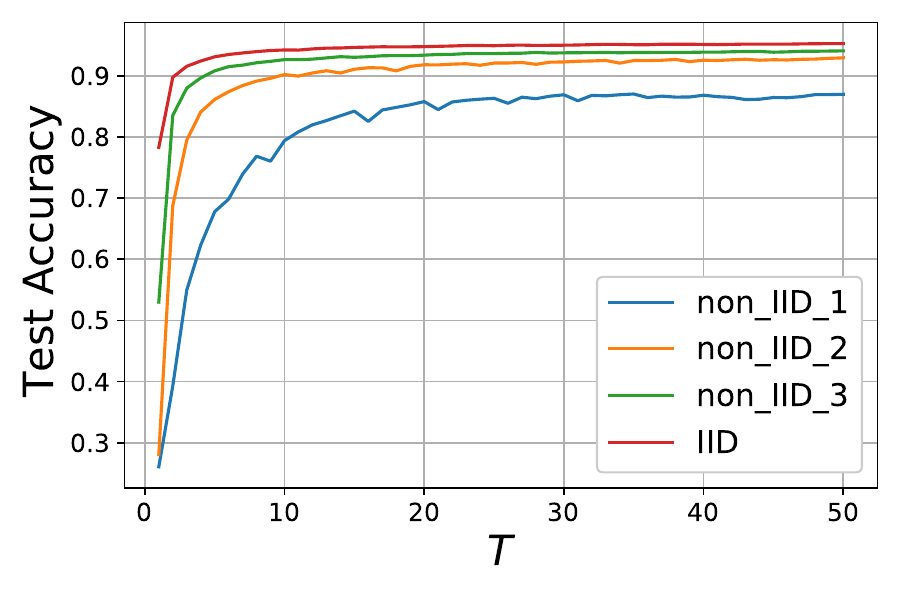}}
\subfigure[Comparison with Inconsistent $J_i$.]{
\includegraphics[width=0.23\textwidth]{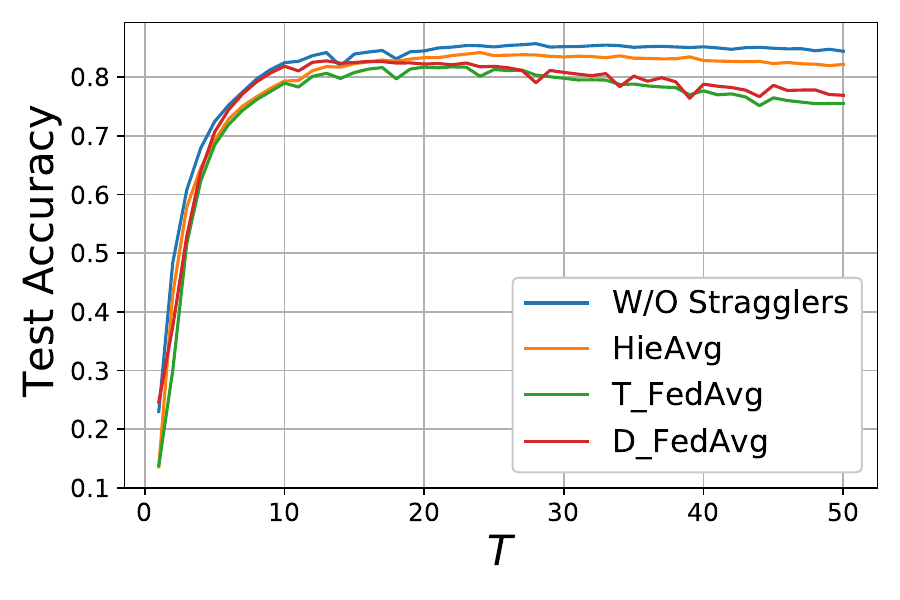}}
\caption{Influences of Data Distribution and Local Device Distribution on BHFL Training.}
\label{fig_k}
\end{figure}

We then investigate the performance of HieAvg with more heterogeneity involved, including different data distributions and inconsistent numbers of local devices at edge servers. For varying data distributions, We adjust the number of image classes in MNIST that each local device holds. For example, non\_IID\_1 means that each local device has at most 1 class of images. The results are presented in Fig. \ref{fig_k}(a), which indicates that when the data distribution is more unbalanced, the model performance is worse. For inconsistent numbers of local devices connected to each edge server, we aim to test the aggregation effectiveness of HieAvg. The results in Fig. \ref{fig_k}(b) show that the BHFL framework with HieAvg can still achieve better performance than the benchmark algorithms. 

\begin{figure}[h]
\centering
\subfigure[Accuracy with Different Algorithms.]{
\includegraphics[width=0.23\textwidth]{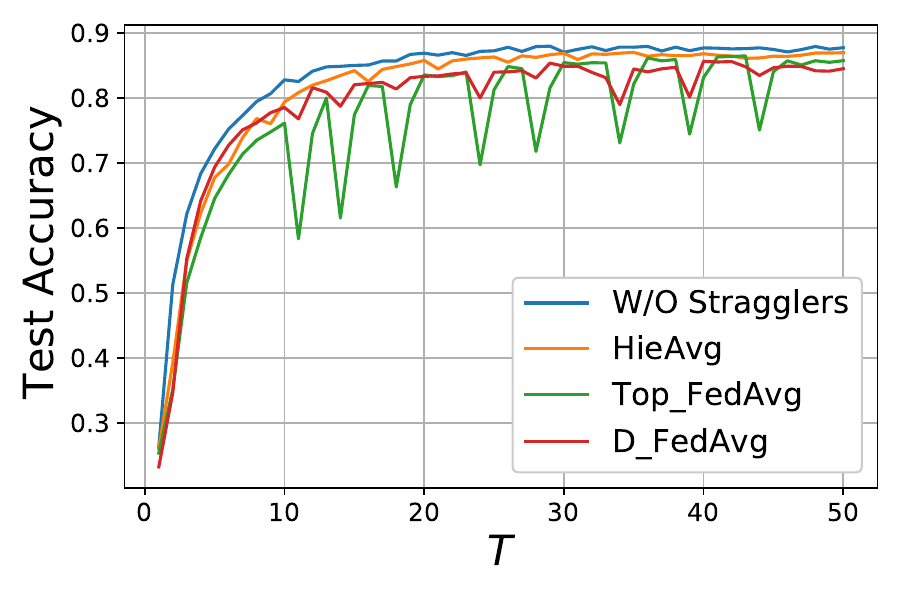}}
\subfigure[Accuracy with Different $J$.]{
\includegraphics[width=0.23\textwidth]{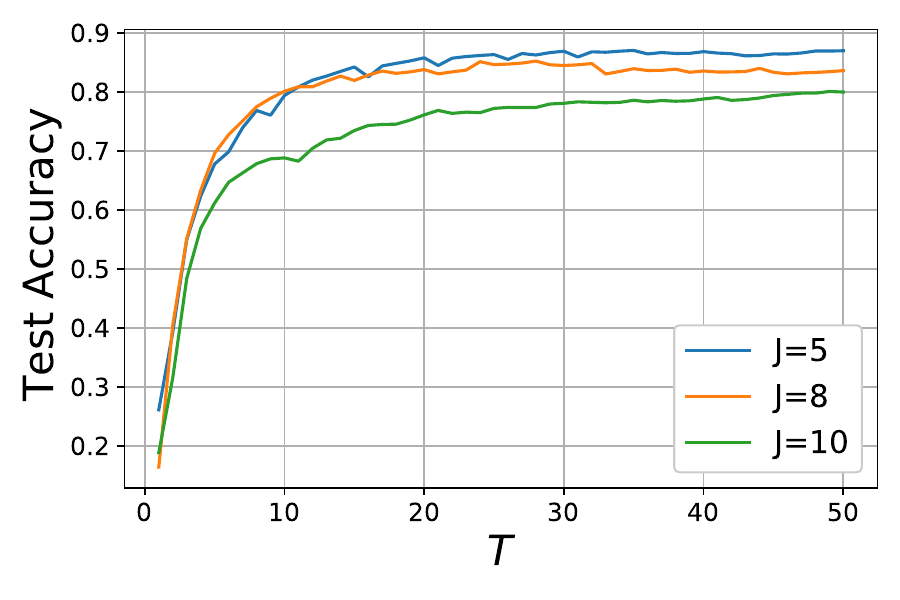}}
\subfigure[Accuracy with Different $N$.]{
\includegraphics[width=0.23\textwidth]{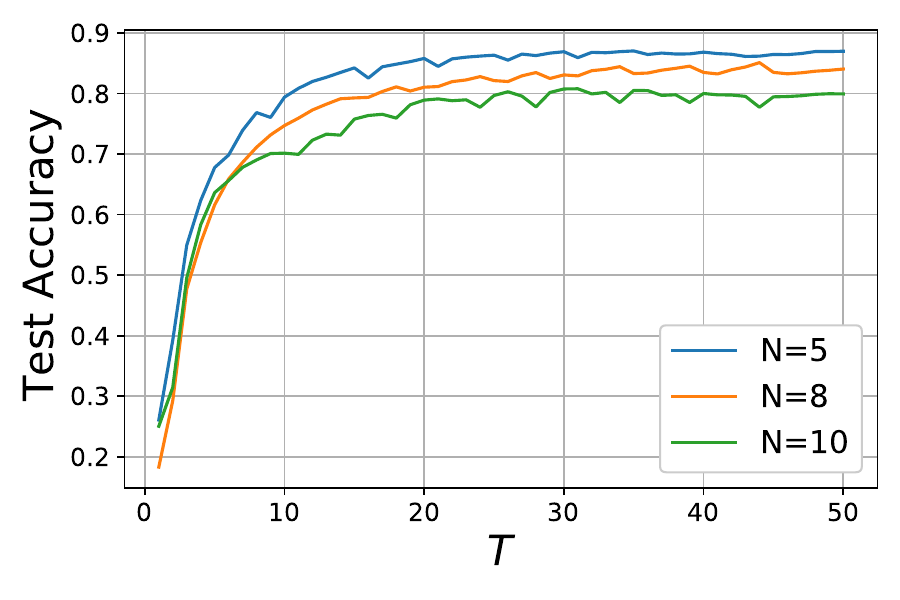}}
\subfigure[Accuracy with Different $K$.]{
\includegraphics[width=0.23\textwidth]{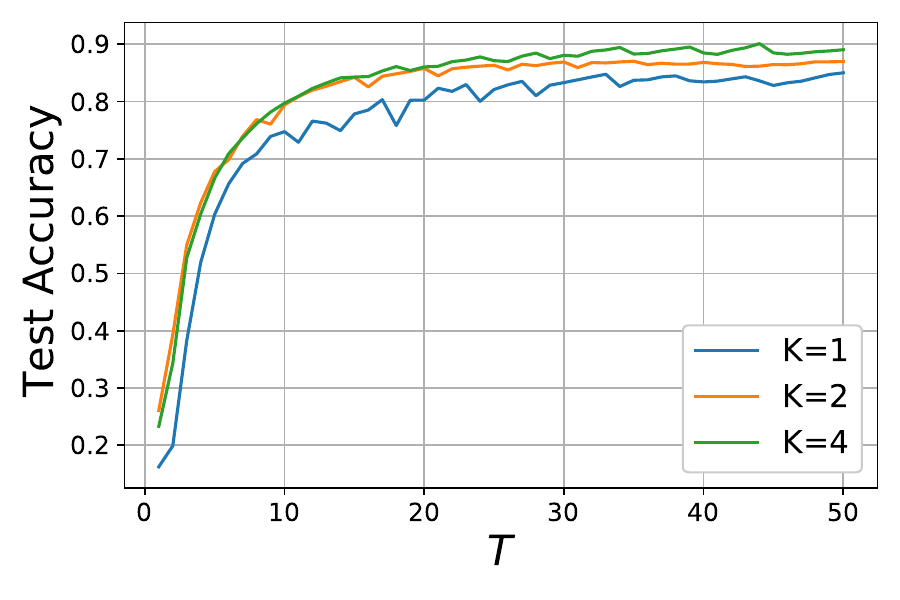}}
\caption{Test Accuracy with Only  Local Device Stragglers.}
\label{fig_local}
\end{figure}

\begin{figure}[h]
\centering
\subfigure[Accuracy with Different Algorithms.]{
\includegraphics[width=0.23\textwidth]{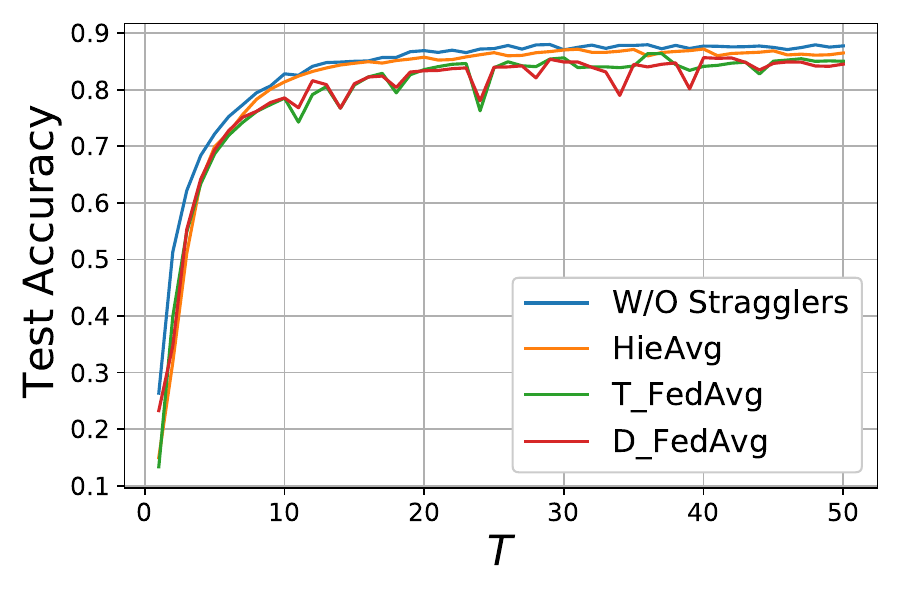}}
\subfigure[Accuracy with Different $J$.]{
\includegraphics[width=0.23\textwidth]{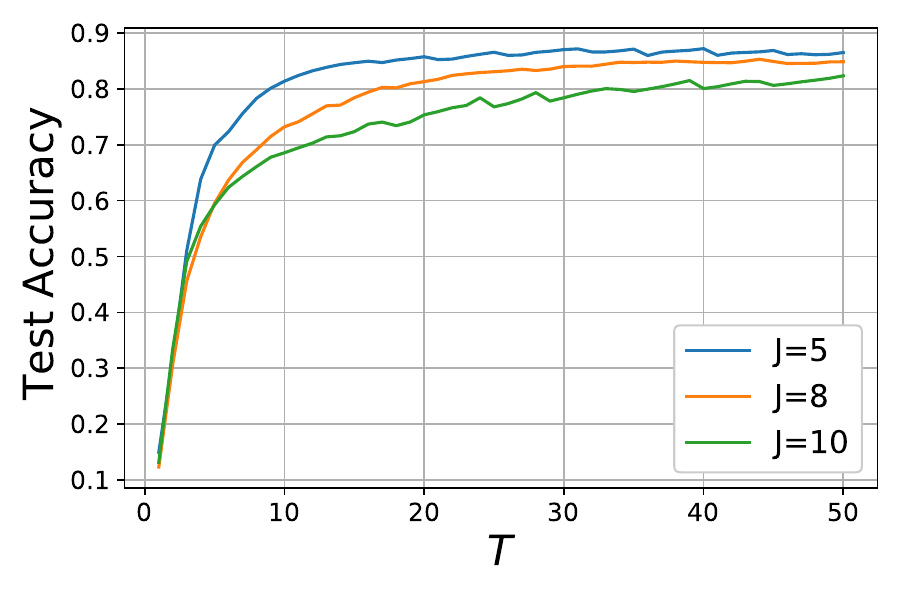}}
\subfigure[Accuracy with Different $N$.]{
\includegraphics[width=0.23\textwidth]{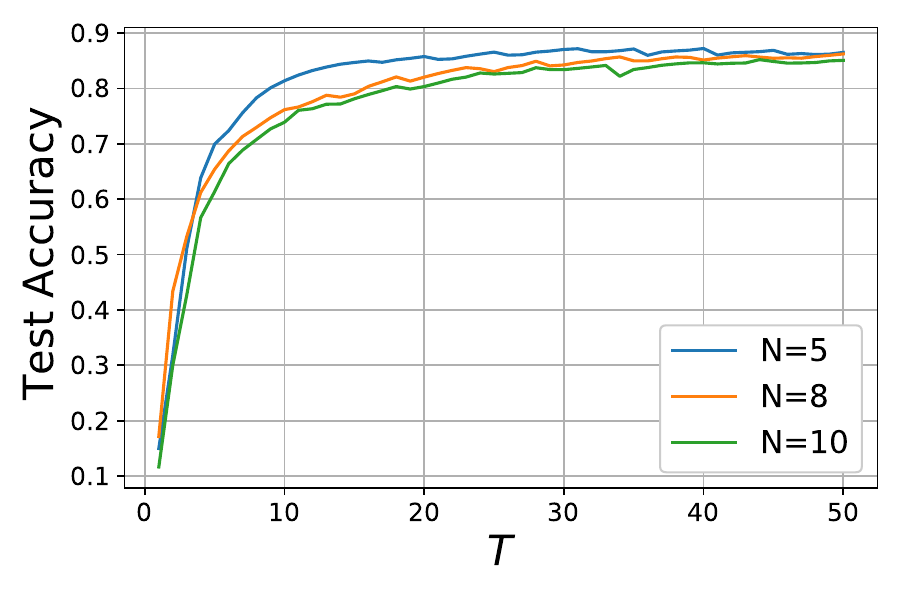}}
\subfigure[Accuracy with Different $K$.]{
\includegraphics[width=0.23\textwidth]{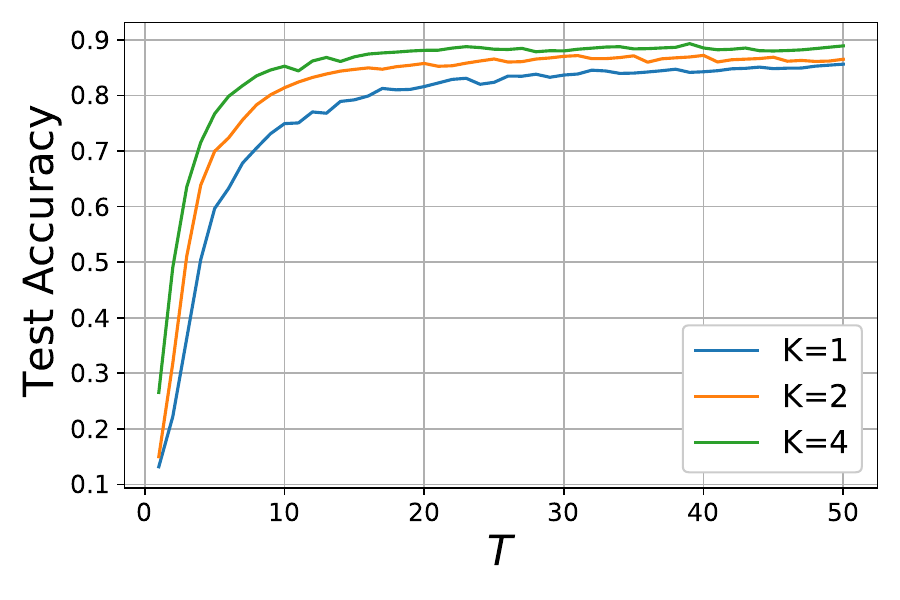}}
\caption{Test Accuracy with Only Edge Server Stragglers.}
\label{fig_edge}
\end{figure}

Finally, we test the convergence performance of HieAvg in mitigating temporary stragglers in only one layer, i.e., the local devices or edge servers. From Fig. \ref{fig_local}(a) and Fig. \ref{fig_edge}(a), we can see HieAvg performs better in both cases compared to other algorithms. Furthermore, by varying the values of $J$, $N$, and $K$, we can find that smaller $J$ and $N$, as well as larger $K$ result in higher accuracy,  
which is consistent with the results in Fig. \ref{fig_h}. These experimental results indicate that HieAvg can also efficiently handle only local device stragglers or only edge server stragglers.

\subsubsection{Evaluation of Latency}
\begin{figure}[h]
\centering
\subfigure[Latency vs. Data Size.]{
\includegraphics[width=0.23\textwidth]{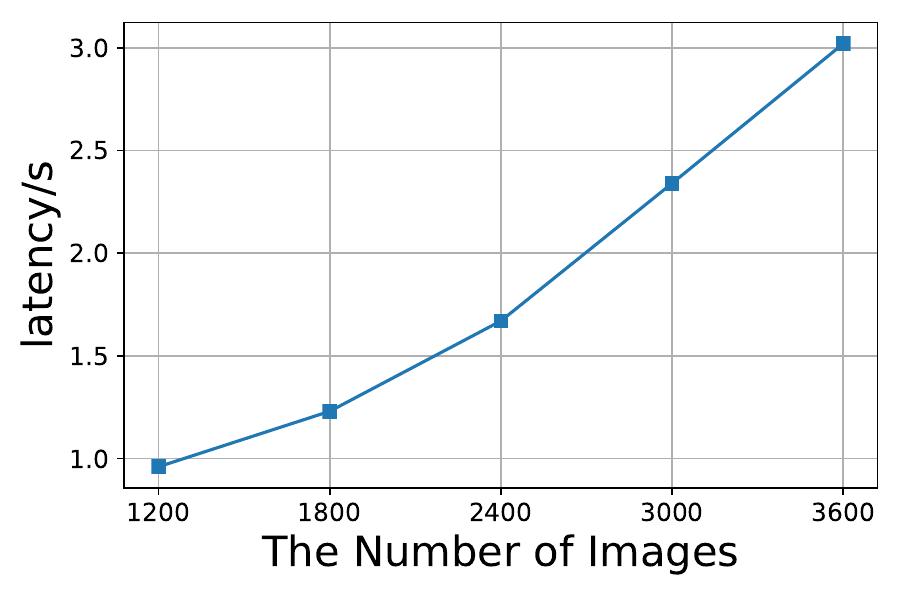}}
\subfigure[$K^*$  vs. Consensus Latency]{
\includegraphics[width=0.23\textwidth]{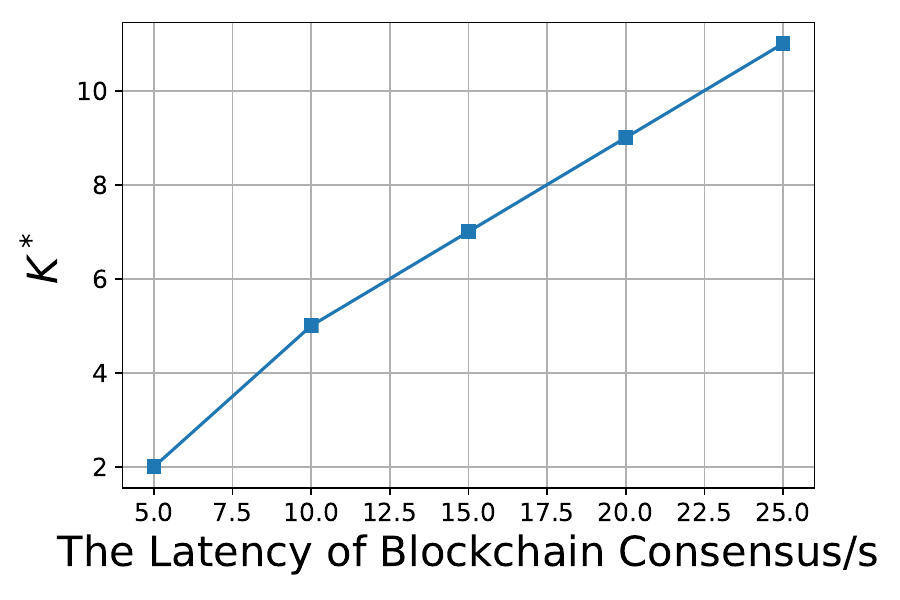}}
\caption{Evaluation Results of Latency Optimization.}
\label{fig_f}
\end{figure}

First, we calculate the computing latency and communication latency on Raspberry Pi and EC2, and the averaged results of three Raspberry Pis are shown in Fig. \ref{fig_f}(a). We can see that the more images on a local device, the higher the latency. This is because the time spent to process more data samples in the local training process will increase. In our basic setting, there are five edge servers and five local devices for each server, so each local device has 2,400 images, and thus, the corresponding latency is around 1.67s. The size of our employed CNN model updates is about 20KB, and the averaged transmission time between Raspberry Pi and EC2 is about 0.51s in the ideal scenario. Inspired by \cite{liu2020client}, we can assume that the latency among edge servers is 0.05s. These parameters are used to solve the latency optimization problem. The latency of Raft-based blockchain consensus will directly influence the optimal value of $K^*$, and the detailed results are illustrated in Fig. \ref{fig_f}(b), which shows that the longer the consensus latency, the more the optimal edge aggregation rounds. Thus, we may adjust $K$ to offset the influence of blockchain consensus latency on the overall latency.

\section{Related Work}\label{related}

Recently, there is an increasing number of studies on HFL. 
Lim \textit{et al.} \cite{lim2021decentralized} propose an HFL framework to reduce node failures and device dropout, and design the resource allocation and incentive mechanisms to improve the learning efficiency  based on game theory. 
Liu \textit{et al.}\cite{liu2020client} propose a client-edge-cloud HFL framework running with the HierFAVG aggregation algorithm and demonstrate that communication efficiency can be improved by introducing the hierarchical architecture in FL. Wang \textit{et al.}\cite{wang2022demystifying} provide theoretical analysis about the convergence of HFL based on Stochastic Gradient Decent (SGD) and emphasize the importance of local aggregation before global aggregation. In \cite{wainakh2020enhancing}, the focus is on protecting participants' privacy in HFL with flexible and  decentralized control. 


With the emergence of blockchain technology, researchers propose the blockchain-based federated learning framework to address the challenges of FL, such as the single point of failure, incentive, and privacy preservation \cite{kim2018device,nguyen2021federated}. There are also some studies applying blockchain in HFL. 
In \cite{xu2022mudfl}, HFL participants are fragmented into multiple microchains to guarantee security and privacy for large-scale IoT intelligence. In \cite{zhang2021bc}, blockchain is used to verify the model updates from edge servers. Nguyen \textit{et al.}  \cite{nguyen2022latency} design a resource allocation mechanism among local devices to assist the latency optimization of BHFL.
As for the challenges of stragglers, we can classify the related research into two categories: coded federated learning (CFL)-based and delayed gradient-based. CFL is proposed in \cite{dhakal2019coded} to speed up FL running the linear regression task, where the basic idea is that local clients transmit the generated coded data to the central server at the beginning of training and the server can compute the coded gradients to compensate the missing gradients of stragglers. In \cite{prakash2020coded}, CodedFedL is designed to mitigate the impact of stragglers on FL that executes linear and non-linear regression. Although CFL performs well in tolerating FL stragglers, it requires extra data transmission and computing, leading to the risk of data privacy leakage and excessive resource consumption. Besides, most CFL-based methods are model-dependent and thus they cannot be generalized to varying deep learning models. As for the method of delayed gradient, AD-SGD is proposed to minimize the difference between the delayed and optimal gradients in \cite{li2021stragglers}. Xu \textit{et al.} \cite{xu2021live} propose the live gradient compensation method to utilize the one-step delayed gradients. However, this kind of method can only deal with the case of stragglers with poor computing power where their partially-trained gradients are still available to the aggregator; while if the stragglers are caused by network connection problems, the server may not get any model updates from the stragglers in that round. 
Recently, a memory enhancement approach, named MIFA, is proposed in \cite{gu2021fast} to solve the problem of stragglers by using stragglers' recently submitted updates to correct their missing updates;
but this approach can lead to significant estimation bias since it only relies on the latest updates which cannot accurately reflect the overall optimization trend.

To overcome the above shortcomings in the existing methods, we propose a novel aggregation method, HieAvg. It can be easily applied to more common cases of FL with non-IID data and even non-convex loss functions 
to solve the problem of stragglers 
in a cost-efficient manner.

\section{Conclusion}\label{conc}
In this paper, we propose a decentralized BHFL framework and design a novel aggregation algorithm HieAvg to ensure the convergence of BHFL even when there are stragglers in both local devices and edge servers, where the data is non-IID and the loss function can be non-convex. We also optimize the overall latency of  BHFL  by jointly considering the requirement of global model convergence and blockchain consensus latency.  Theoretical analysis for the convergence of HieAvg is provided and extensive experiments are conducted to demonstrate the validity and superiority of our proposed schemes. 
In the future, we would like to design incentive and privacy protection mechanisms to further improve the performance of BHFL.



\bibliographystyle{IEEEtran}
\bibliography{main.bib}

\begin{IEEEbiography}[{\includegraphics[width=1in,height=1.25in,clip,keepaspectratio]{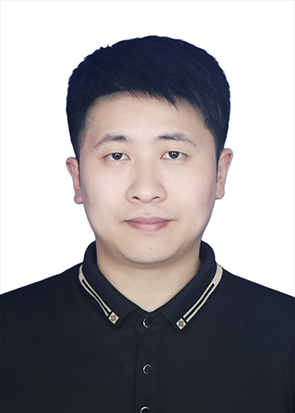}}]{Zhilin Wang} received his B.S. from Nanchang University in 2020. He is currently pursuing his Ph.D. degree in Computer and Information Science at Indiana University-Purdue University Indianapolis (IUPUI). He is a Research Assistant with IUPUI, and he is the reviewer of IEEE TPDS, IEEE IoTJ, Elsevier JNCA, IEEE TCCN, IEEE ICC'22, IEEE Access, and Elsevier HCC; he also serves as the TPC member of the IEEE ICC'22 Workshop. His research interests include blockchain, federated learning, edge computing, and optimization theory.
\end{IEEEbiography}

\begin{IEEEbiography}[{\includegraphics[width=1in,height=1.25in,clip,keepaspectratio]{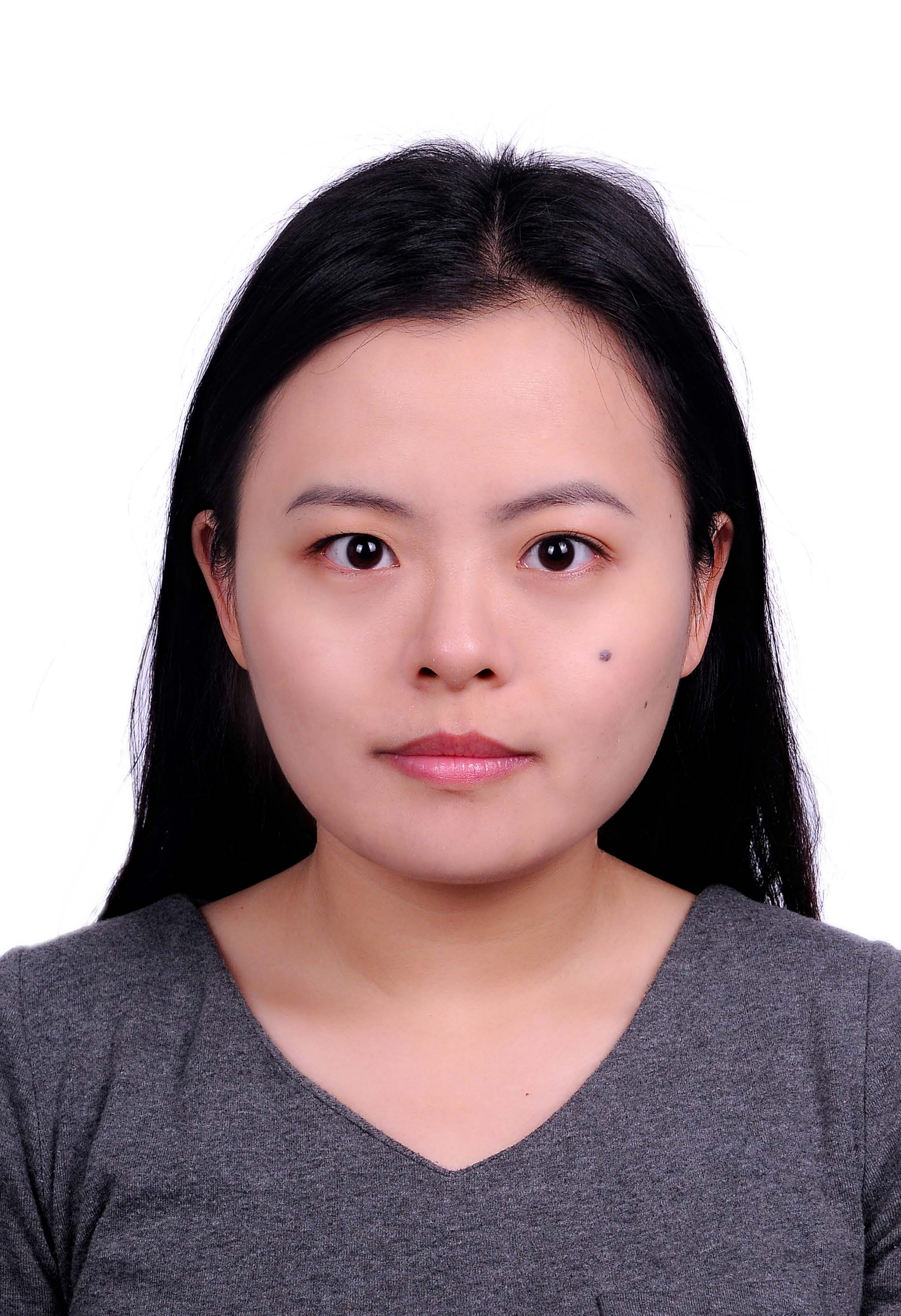}}]{Qin Hu} received her Ph.D. degree in Computer Science from the George Washington University in 2019. She is currently an Assistant Professor with the Department of Computer and Information Science, Indiana University-Purdue University Indianapolis (IUPUI). She has served on the Editorial Board of two journals, the Guest Editor for multiple journals, the TPC/Publicity Co-chair for several workshops, and the TPC Member for several international conferences. Her research interests include wireless and mobile security, edge computing, blockchain, and federated learning.
\end{IEEEbiography}

\begin{IEEEbiography}[{\includegraphics[width=1in,height=1.25in,clip,keepaspectratio]{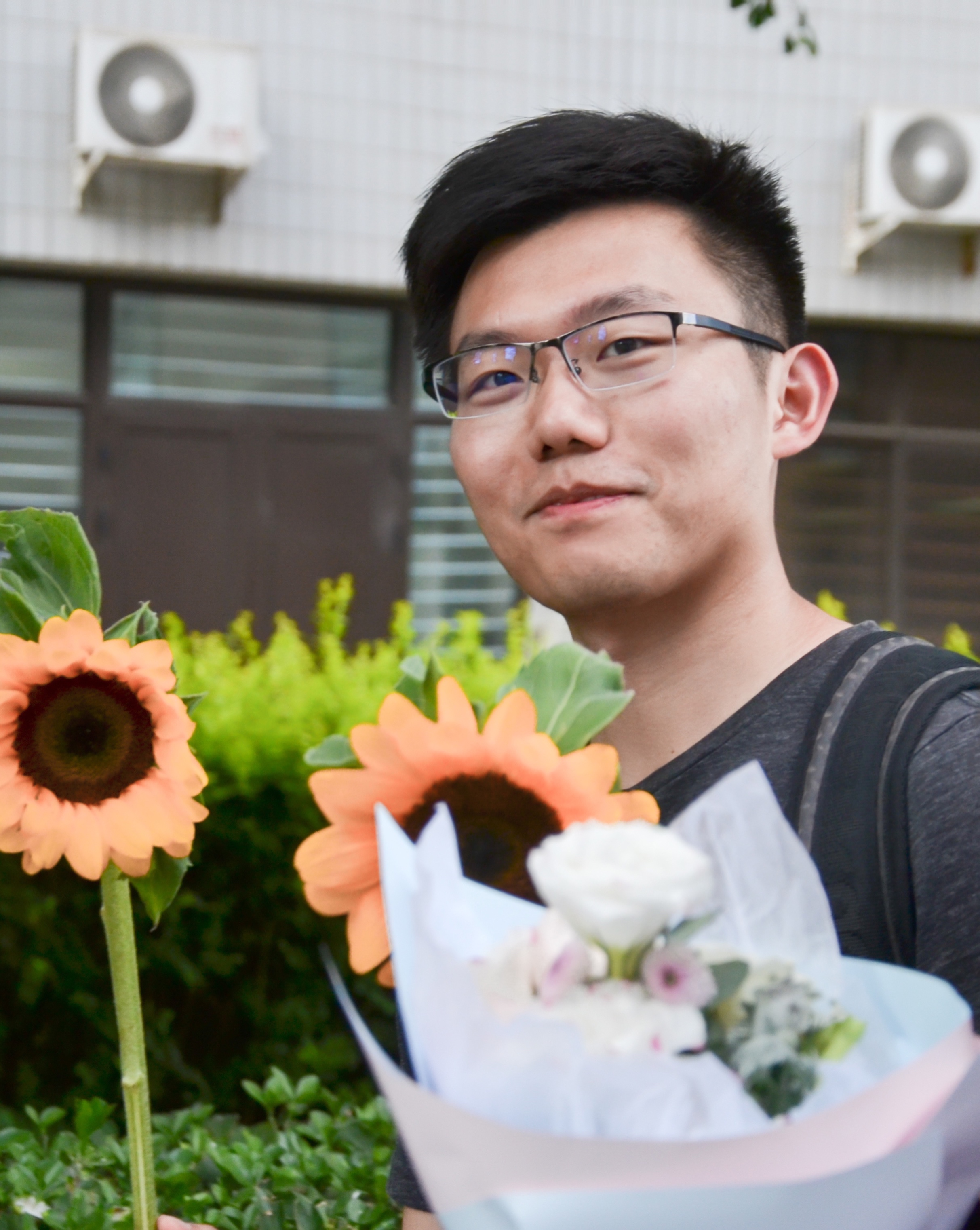}}]{Minghui Xu} received the BS degree in Physics from the Beijing Normal University, Beijing, China, in 2018, and the PhD degree in Computer Science from The George Washington University, Washington DC, USA, in 2021. He is currently an Assistant Professor in the School of Computer Science and Technology, Shandong University, China. His current research focuses on blockchain, distributed computing,  and quantum computing.
\end{IEEEbiography}

\begin{IEEEbiography}[{\includegraphics[width=1in,height=1.25in,clip,keepaspectratio]{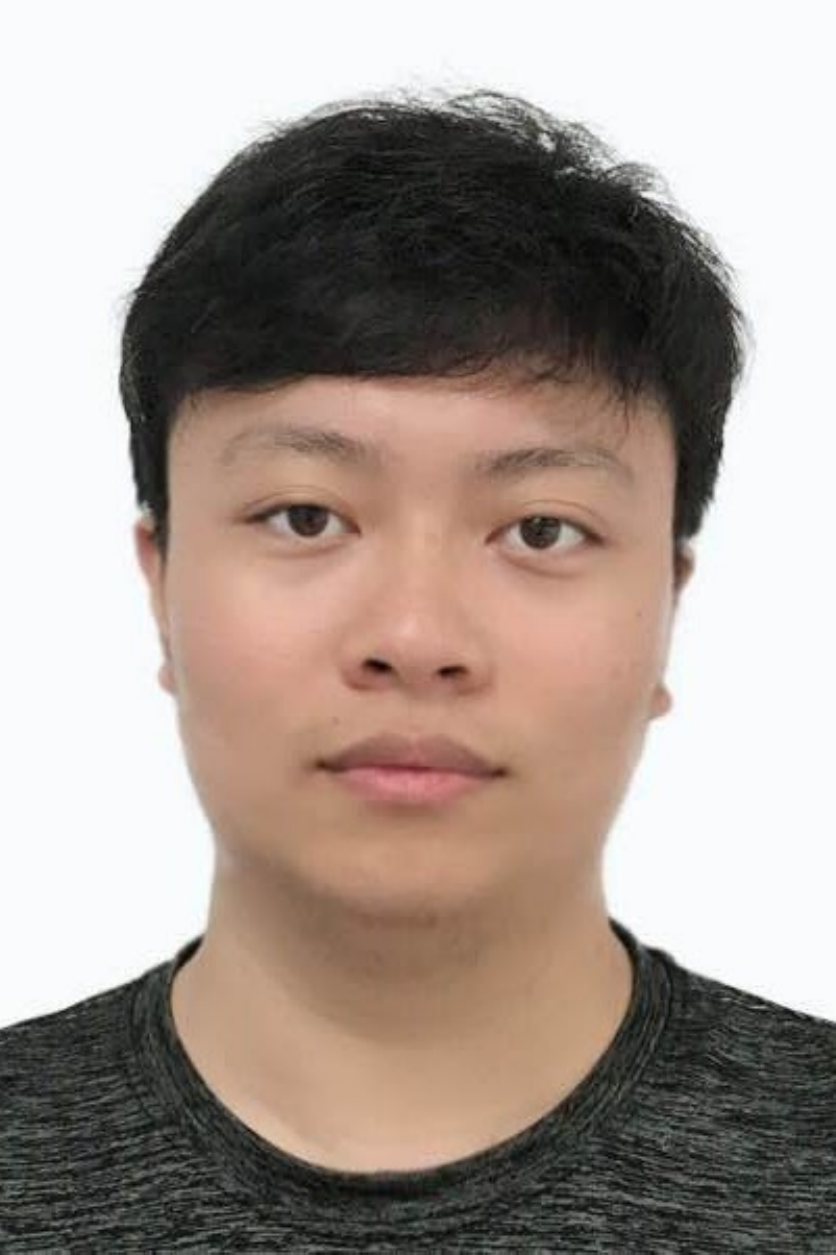}}]{Zehui Xiong} is currently an Assistant Professor in the Pillar of Information Systems Technology and Design, Singapore University of Technology and Design. Prior to that, he was a researcher with Alibaba-NTU Joint Research Institute, Singapore. He received the PhD degree in Nanyang Technological University, Singapore. He was the visiting scholar at Princeton University and University of Waterloo. His research interests include wireless communications, network games and economics, blockchain, and edge intelligence. He has published more than 140 research papers in leading journals and flagship conferences and many of them are ESI Highly Cited Papers. He has won over 10 Best Paper Awards in international conferences and is listed in the World’s Top $2\%$ Scientists identified by Stanford University. He is now serving as the editor or guest editor for many leading journals including IEEE JSAC, TVT, IoTJ, TCCN, TNSE, ISJ, JAS.
\end{IEEEbiography}

\newpage

\end{document}